\title{Quantifying neural network uncertainty under volatility clustering}
\author{
        Steven Y. K. Wong\thanks{Steven Wong and Jennifer Chan are with School of Mathematics and Statistics, University of Sydney, Australia. Corresponding email: steven.ykwong87@gmail.com.},
        Jennifer S. K. Chan\footnotemark[1],
        Lamiae Azizi\thanks{Lamiae Azizi is with NABLAS AI, Australia.}
}
\date{\today}
\documentclass[12pt]{article}

\usepackage[utf8]{inputenc}

\usepackage{amsmath,amssymb,amsthm}
\usepackage{graphicx}

\usepackage{longtable}

\usepackage{booktabs}
\usepackage{siunitx}
\usepackage{natbib}

\usepackage{hyperref}
\usepackage{url}
\usepackage{bm}
\usepackage{float}
\usepackage{flafter}
\usepackage{textcomp}
\usepackage[super]{nth}     
\usepackage{nicefrac}       
\usepackage{makecell}
\usepackage{adjustbox}
\usepackage{subcaption}
\usepackage{mathtools}      

\newcommand{\code}[1]{\texttt{#1}}
\DeclareMathOperator*{\argmin}{\arg\min}

\newcommand{\Ell}{\mathcal{L}}

\newcommand{\D}{\mathcal{D}}
\newcommand{\Out}{\mathrm{O}}

\newcommand*\diff{\mathop{}\!\mathrm{d}}
\newcommand*\prob{\mathrm{p}}

\DeclareMathOperator{\Nat}{\mathbb{N}}
\DeclareMathOperator{\R}{\mathbb{R}}
\DeclareMathOperator{\Z}{\mathbb{Z}}

\DeclareMathOperator{\N}{N}
\DeclareMathOperator{\G}{G}
\DeclareMathOperator{\IG}{IG}
\DeclareMathOperator{\Gam}{Gamma}

\DeclareMathOperator{\tdist}{St}
\DeclareMathOperator{\NormInvGam}{NIG}
\DeclareMathOperator{\SMD}{SMD}
\DeclareMathOperator{\E}{E}
\DeclareMathOperator{\Var}{Var}

\newcommand{\ignore}[1]{}{}

\begin{document}
\maketitle

\begin{abstract}
    Time-series with volatility clustering pose a unique challenge to uncertainty quantification (UQ) for returns forecasts.
    Methods for UQ such as Deep Evidential regression offer a simple way of quantifying return forecast uncertainty without the costs of a full Bayesian treatment.
    However, the Normal-Inverse-Gamma (NIG) prior adopted by Deep Evidential regression is prone to miscalibration as the NIG prior is assigned to latent mean and variance parameters in a hierarchical structure.
    Moreover, it also overparameterizes the marginal data distribution. These limitations may affect the accurate delineation of epistemic (model) and aleatoric (data) uncertainties. 
    We propose a Scale Mixture Distribution as a simpler alternative which can provide favorable complexity-accuracy trade-off and assign separate subnetworks to each model parameter.   
    To illustrate the performance of our proposed method, we apply it to two sets of financial time-series exhibiting volatility clustering: cryptocurrencies and U.S. equities and test the performance in some ablation studies.
\end{abstract}

\section{Introduction} \label{sec:unc_intro}

\subsection{Motivation} \label{sec:motivation}
In finance, volatility estimation is essential for investors, traders, and financial institutions for risk management serving as a crucial metric for assessing and predicting the degree of price fluctuations in assets. Forecast uncertainty is an important determinant of the optimal trade-off between return and risk, and has long been used in portfolio construction, such as in Kelly criterion \citep{KellyCriterion,Byrnes:2018} and Bayesian-based portfolio optimization \citep{BlackLitterman:1991}.

Recently, neural network emerged as an optimal method for diverse regression problems due to its remarkable capacity for linear and non-linear mapping, surpassing the constraints of rigid model structures in traditional statistical models. In this paper, we are concerned with neural network UQ in financial time-series that exhibit \emph{volatility clustering} \citep{Cont:2001} --- irregular bursts of high volatility that cluster in time.
This poses a challenge to practitioners wishing to use neural networks to forecast both asset returns and uncertainties, as irregular bursts of volatility can induce forecast errors.
Moreover, challenges also stem from modelling these irregular bursts of volatility using \emph{classical} neural networks which are typically trained using mean squared error (MSE) and provide point estimates for the target variable $Y$ using the mean prediction $\mu$ conditional on the input $X$ taking value $x$.
As MSE considers only the difference of the predicted mean $\mu$ from the observed value of $Y$, it is incapable of estimating the conditional variance $\sigma^2$  which refers to the uncertainty of the noisy data $Y$ in UQ (\citealp{Goodfellow:2016}). 

\subsection{Uncertainty quantification}

This conditional variance $\sigma^2$ originates from the stochastic relationship between the input variable $X$ and output variable $Y$ \citep{SourcesOfUncertainty:2023}.
It is termed \emph{aleatoric} uncertainty --- uncertainty in the data that is irreducible through additional information (i.e., more observations).
As long as the distribution of $Y|x$ is not degenerate (i.e., $Y$ cannot be perfectly predicted using $X$), there will always be aleatoric uncertainty and is indicated by the variance of $Y|x$.
It can be estimated by assigning a distribution that is most likely to have generated the data $Y|x$.


In addition to aleatoric uncertainty, there is also uncertainty attributable to the model.
This source of uncertainty is known as \emph{epistemic} uncertainty.
Epistemic uncertainty can be further decomposed into \emph{model} uncertainty, which relates to the correct specification of the model, such as assumption of normality or linear relationship betwen $Y$ and $X$, and \emph{parametric} uncertainty, which relates to the correct estimation of model parameters such as mean $\mu$ and variance $\sigma^2$ \citep{Sullivan:2015,SourcesOfUncertainty:2023}.
Epistemic uncertainty is \emph{reducible} through additional information (e.g., more observations and additional variables) and typically scales inversely with sample size \citep{Meinert:2022}.
In practice, a clear separation between aleatoric and epistemic uncertainties is often impossible.
To illustrate, consider the (fair) dice rolling experiment, commonly considered to be a process of pure randomness.
However, if the initial position and each rotation of the dice can be measured, then it is possible to predict the outcome of each dice roll \citep{Hora:1996,SourcesOfUncertainty:2023}.
Thus, what is truly aleatoric (i.e., unpredictability of dice roll) and what is epistemic (i.e., initial position and rotation of the dice are merely missing variables) may be difficult to disentangle from a philosophical perspective.

As a modeller (in our case, an investor), one is concerned with \emph{predictive uncertainty} \citep{Gawlikowski:2021} which is the total uncertainty around a point estimate.
Predictive uncertainty is comprised of \emph{aleatoric} uncertainty, and \emph{epistemic} uncertainty \citep{SourcesOfUncertainty:2023}.
Together, these two sources of uncertainty give a complete picture of the uncertainty of the forecasts\footnote{We provide a more in-depth discussion of the types of uncertainty in Sections~\ref{sec:nig_prior} and \ref{sec:unc_scale_mixture}.}.
To account for both sources of uncertainty, a full Bayesian approach not only assumes a data distribution but also assigns prior distributions to all model parameters.
In contrast, the classical likelihood approach considers only the data distribution, often leading to criticisms for underestimating the true level of uncertainty.
Similarly, the Bayesian neural network (BNN), where prior distributions are placed on network weights, are able to capture both sources of uncertainty. 
Simpler BNNs can be trained using Markov Chain Monte Carlo (MCMC) which offers accurate estimates of the posterior distribution, while more complex BNNs can be trained using variational inference which trades off accuracy for tractability \citep{Jospin:2022}.
However, a full BNN is often both computationally costly and requires modification of the network architecture.

\subsection{Research background}

Recent advances (see \citealp{Gawlikowski:2021} for a recent survey) focuses on predicting the conditional distribution of $Y|x$  by estimating the parameters of the distribution, including both mean $\mu$ and variance $\sigma^2$ using the negative log-likelihood (NLL) loss function instead of MSE.
This type of neural network is called \emph{density network} \citep{Bishop:1994} and provides an attractive alternative to BNN, as it allows priors to be placed on parameters of the assumed data distribution.
Evaluation of the marginal distribution NLL is done by integrating out the prior (i.e., uncertainty of the model parameters), in the spirit of posterior sampling using MCMC for BNNs.
Hence, this approach offers an attractive trade-off between adequately quantifying uncertainty as well as avoiding the computational cost of a full Bayesian treatment.
Prominent methods in this category include \emph{Deep Ensemble} (\citealp{Lakshminarayanan:2017}; the \emph{Ensemble} method) and \emph{Deep Evidential} (\citealp{DeepEvidentialNet:2020}; the \emph{Evidential} method).
Between the two methods, Ensemble assumes a Normal data distribution with mean and variance parameters, $(\mu, \sigma^2)$, where $\sigma^2$ indicates aleatoric uncertainty.
Epistemic uncertainty is estimated by using an ensemble of randomly initialized networks, where each network settles at a different local minima and thus giving different mean and variance estimates.
Hence, these estimates reveal parameter uncertainty.

By contrast, Evidential improves on this by assuming a Normal prior for the mean parameter $\mu$ and an Inverse-Gamma prior for the variance parameter $\sigma^2$ of the normal data distribution.
This is known as the Normal-Inverse-Gamma (NIG) prior.
The resultant marginal distribution after integrating out the priors of $\mu$ and $\sigma^2$ is a Student's t-distribution parameterized by parameters of the NIG priors \citep{BayesianTheory}.
This Student's t-distribution is a hierarchical distribution consisting of the Normal data distribution conditional on parameters $\mu, \sigma^2$, and the prior distribution for each of $\mu$ and $\sigma^2$. 
As t-distribution has heavier tails than Normal distribution, this hierarchical distribution provides a principled way of capturing epistemic uncertainty without the need for ensembling.
Moreover, Evidential requires only minimal modifications to the network architecture, specifically a new output layer and training with the NLL of the marginal distribution.

However, \cite{Bengs:2023} argued that hierarchical models, such as Evidential, cannot be well calibrated as the prior distributions are assigned to the latent model parameters $\mu, \sigma^2$ which  cannot be observed directly.
Thus, hierarchical models lack theoretical guarantee on the robustness of their estimated distributions. 
Moreover, \cite{Meinert:2022} argued that the marginal Student-t distribution of the Evidential method is overparameterized.
A detailed explanation is provided in Section~\ref{sec:shortcomings}.
Hence, epistemic uncertainty is difficult to estimate objectively.
In addition to the shortcomings from a statistical modelling perspective, we also note an architectural shortcoming.
In both Ensemble and Evidential methods, output of the final hidden layer is fed into a linear output layer, which outputs two or four hyperparameters of the Normal or marginal t-distribution\footnote{In the case of Evidential's marginal t-distribution, softplus is applied to the output for parameters $\nu, \alpha, \beta$.}, respectively.
We consider this a weakness of these approaches as the latent representation has to provide a sufficiently rich encoding to linearly derive all hyperparameters of the distribution.

Recent works have extended Evidential into various related areas.
\cite{Huttel:2023} extended Evidential into nonparametric quantile regressions by replacing the Normal data distribution with an Asymmetric Laplace (AL) distribution with NIG prior.
By expressing AL into a mean-variance mixture of Normal distribution and an exponentially distributed mixing variable $z_i$, they obtained a marginal t-distribution conditional on $z_i$ that is similar to the Evidential method.
Other works have focused on regions where the neural network is uncertain due to the lack of training observations, called \emph{high uncertainty areas}.
In Evidential, \cite{DeepEvidentialNet:2020} proposed to regularize the network with a \emph{total evidence} penalty added to the marginal NLL to encourage the network to output lower evidence ($\nu$, to be introduced in Section~\ref{sec:deep_evidential}), and thus greater uncertainty.
In this high uncertainty region, \cite{Ye:2024} argued that the gradient of the marginal NLL of Evidential is close to zero during training when $\alpha \to 1$ (a parameter of the NIG prior).
They proposed to add another regularization term in addition to the total evidence penalty to address the zero gradient problem and showed that the proposed regularization achieved comparable performance to Evidential on the University of California Irvine Machine Learning Repository (UCI) benchmark dataset.
Similarly focusing on sparse training samples, \cite{Deng:2023} proposed to use the Fisher Information Matrix to up-weight uncertain observations for classification problems.
Separately, \cite{MeinertLavin:2022} proposed the \emph{Multivariate Deep Evidential regression}, using a Normal-Inverse-Wishart prior as a multivariate extension of NIG prior for multivariate regressions.
Importantly, \cite{MeinertLavin:2022} argued that the total evidence penalty is only necessary as the NIG prior is overparameterized.
As a workaround, \cite{MeinertLavin:2022} suggested to fix $\nu = c\beta$ to reduce the number of parameters of the NIG prior from four to three, where $c$ is a hyperparameter of the model\footnote{\cite{MeinertLavin:2022} used notations $\nu = r\kappa$, where $\kappa$ maps to $\beta$ in our notations.}.
While these advances extend the statistical model framework and address some shortcomings of Evidential, we propose an alternative formulation with a simplified statistical model structure which alleviates the need for a total evidence penalty, while also offering potential research directions to nonparametric quantile regression and multivariate regression.

\subsection{Proposed method} \label{sec:proposed_method}

To tackle the shortcomings highlighted by \cite{Meinert:2022},  \cite{Bengs:2023} and others, we combine and extend the ideas of Ensemble and Evidential methods into a framework called the \emph{Combined} method for quantifying forecast uncertainty of financial time-series. Specifically, 
we propose to formulate the data distribution using the \emph{scale mixture distribution} (SMD), a simpler alternative to the NIG prior, by avoiding prior distributions for the latent mean and variance parameters of the Normal distribution.
Specifically, the epistemic uncertainty in SMD is captured by allowing each normally distributed data point to have its own variance $\sigma^2 \nu^{-1}$, where $\nu$ is the \emph{scaling} factor.
This scaling factor provides flexibility to each data point by enlarging or diminishing the variance for outliers or modal points.
Then, a sole Gamma prior is assigned to the scaling factor.
Integrating out the scaling factor of SMD also results in a marginal t-distribution and its variance indicates predictive uncertainty, which stems from two sources of uncertainty --- the aleatoric uncertainty from the $\sigma^2$ of the conditional Normal data distribution and epistemic uncertainty from the Gamma distribution of scaling factor.
This simplification reduces the number of effective parameters by one and resolves the overparameterization of NIG, as highlighted by \cite{Meinert:2022}.
This also alleviates the need to regularize the marginal NLL, as noted by \cite{MeinertLavin:2022}.
A detailed explanation is provided in Section~\ref{sec:address_shortcome}. 

In addition to statistical model structure, we also propose a novel architecture to model parameters of the marginal distribution using disjoint subnetworks, rather than a single output layer as in Ensemble and Evidential.
We show through an ablation study in Section~\ref{sec:unc_ablation_study} that the flexibility offered by the subnetworks is crucial to forecasting predictive uncertainty that closely tracks volatility when the time-series exhibit volatility clustering.
As return forecast accuracy is also important in finance, we incorporate model averaging (in the spirit of ensemble) into our Combined method and show that it significantly improves return forecast accuracy without significantly increasing the estimated predictive uncertainty.

To illustrate our contributions, we apply our proposed method to cryptocurrency and U.S. equities time-series forecasting.
Cryptocurrencies are an emerging class of digital assets.
They are highly volatile and frequently exhibit price bubbles \citep{Fry:2016,Hafner:2018,Chen:2019,phillip2018new,Nunez:2019,Petukhina:2021}, with large volumes of high frequency data (e.g., prices in hourly intervals) freely available from major exchanges.
This makes cryptocurrencies an ideal testbed for UQ methodologies in financial applications.
Given the extreme levels of volatility, we view cryptocurrencies as one of the most challenging datasets for this type of application.
A comparison in U.S. equities is also provided which illustrates performance in conventional financial time-series.

\subsection{Contributions and layout} 
A summary of our contributions is as follows:
\begin{itemize}
    \item We propose to place a prior solely on the scaling factor $\nu$ using a SMD, rather than on both mean and variance using the NIG prior.
    In Table~\ref{tab:nig_v_ng} (Appendix~\ref{apd:uci}), we compare SMD to NIG using the same network architecture on the UCI dataset.
    We observe that both RMSE and NLL are overwhelmingly in SMD's favor, achieving superior results in 6 (of 9) and 8 (of 9) of the datasets for RMSE and NLL, respectively.
    We attribute this to the simpler formulation of SMD, which resolves the latent variables issue of hierarchical models noted by \cite{Bengs:2023} and the overparameterization problem of the NIG prior, noted by \cite{Meinert:2022}.
    \item We show that modelling hyperparameters of the posterior distribution using subnetworks rather than a single output layer (as per Ensemble and Evidential) results in superior performance.
    We show in the ablation study (Section~\ref{sec:unc_ablation_study}, the ``Single Output'' model in Table~\ref{tab:ablation} and Figure~\ref{fig:ablation}) that separate modelling of the hyperparameters is essential in producing predictive uncertainty estimates that closely track actual realized forecast error of Bitcoin and Chevron.
    We attribute this to the ease of specialization when each hyperparameter of the posterior distribution is modelled through disjoint subnetworks.
    \item In additional to the subnetwork modelling of distribution hyperparameters, we show that including squared returns as an additional feature to the long short-term memory (LSTM; \citealp{LSTM:1997}) layers in architectural design is also essential to successfully quantifying uncertainty in financial time-series.
    We provide a precedence on UQ in financial time-series, a neglected application in UQ literature.
\end{itemize}
In the rest of this paper, we first describe the setup of our motivating application (asset return forecasting) in Section~\ref{sec:unc_setup} and review related works in Sections~\ref{sec:deep_ensemble} and \ref{sec:deep_evidential}.
We describe and discuss our proposed framework in Section~\ref{sec:combined}.
Data description and empirical results of applying Ensemble, Evidential and Combined methods on cryptocurrency and U.S. equities are detailed in Section~\ref{sec:unc_experiments}.
Whilst this paper is focused on UQ in time-series that exhibit volatility clustering, in Appendix~\ref{apd:uci}, we also provide a direct comparison to Evidential and Ensemble methods using the UCI benchmark datasets (non-time-series), as previously analyzed in \cite{ProbBackprop:2015}, \cite{DropoutBayesian:2016}, \cite{Lakshminarayanan:2017}, and \cite{DeepEvidentialNet:2020}.
Finally, concluding remarks are provided in Section~\ref{sec:unc_conclusion}.

\section{Methodological development in uncertainty quantification}

\subsection{Problem setup} \label{sec:unc_setup}

Consider an investor making iterative forecasts of asset returns.
At every period $t \in \{1, \dots, T\}$, an investor observes price history up to $t$ and uses the preceding $\{K \in \Z | 0 < K < t\}$ period returns to forecast one-step ahead returns.
We define an asset's return at time $t$ as the log difference in price $r_t = \log{p_t} - \log{p_{t-1}}$ and, consistent with empirical findings in finance literature \citep{PesaranTimmermann:1995,Cont:2001}, we assume that the data generation process (DGP) is time-varying:
\begin{equation}
    r_t \sim \N(\mu_t, \sigma_t^2). \label{eq:unc_dgp}
\end{equation}

Let $\bm{\zeta}_t = (\mu_t, \sigma_t^2)$ be parameters of the assumed DGP, $\bm{x}_{t-1} = \{r_{t-K}, r_{t-K+1}, \dots, r_{t-1}\}$ be a $K$-length input sequence\footnote{For illustrative purposes, we state that the sequence only contains returns $r_t$. However, as discussed in Section~\ref{sec:unc_architecture}, we also include squared returns $r_t^2$ as part of the input sequence.} using returns up to $t-1$ and $y_{t-1} = r_t$ be forward one period return.
The training dataset is comprised of $\D_t = \{(\bm{x}_{q-1}, y_{q-1}) | q \in \Nat : q \leq t\}$ input-output pairs\footnote{Note that at each portfolio selection period $t$, the training set can at most contain data up to $t-1$ as we have not yet observed $r_{t+1}$.} and is essentially a set of sequences formed with a $K$-length sliding window and their corresponding regression targets.
Our goal is to forecast $y_t$ (which corresponds to $r_{t+1}$).
At each $t$, the investor's goal is to solve the optimization problem\footnote{For clarity, the case of a single asset is shown. At each $t$, there are $N$ assets and the dataset is typically in a $t \times N$ layout. It is easy to see the generalization of Equation~\eqref{eq:generic_loss} over $N$ assets, where the average loss is calculated over $(t - K - 1) \times N$ instances.},
\begin{equation}
    \bm{\theta}_t = \argmin_{\bm{\theta}^*}-\sum_{q=K}^{t-1}\log \prob\big(y_{q}|F(\bm{x}_{q};\bm{\theta}^*)\big), \label{eq:generic_loss}
\end{equation}
where $F(\bm{x}; \bm{\theta})$ is a neural network with input $\bm{x}$ and parameters $\bm{\theta}$, $\bm{\theta} = \bigcup_{\ell=1}^{L} \{\bm{W}^{(\ell)}, \bm{b}^{(\ell)}\}$ is the set of network weights and biases and   $\prob\big(y|F(\bm{x};\bm{\theta})\big)$ is the likelihood of observing $y$ based on the outputs of neural network $F(\cdot; \cdot)$ and the assumed marginal distribution.
In other words, the investor is concerned with recovering the parameters $\hat{\bm{\zeta}}_t = (\hat{\mu}_t, \hat{\sigma}_t^2) \coloneqq F(\bm{x}_t;\bm{\theta}_t)$ that are most likely to have generated the observed data.
In this setup, $\hat{\sigma}_t^2$ can be interpreted as an estimate of aleatoric uncertainty which is the contemporaneous variance of the DGP at time $t$.

There are two parts to this problem.
The first part concerns UQ specifically for time-series that exhibit volatility clustering and is the primary focus of this work.
The second part concerns advancing methods of UQ across general applications.
In Appendix~\ref{apd:uci}, we show that our proposed approach can still benefit non-time-series problems in spite of it being designed to deal with a series of data points indexed in time order and exhibiting volatility clustering.

\subsection{Deep Ensemble method} \label{sec:deep_ensemble}


\cite{Lakshminarayanan:2017} proposed the Ensemble method which assumes that the regression target $y$ is drawn from a Normal distribution with mean $\mu$ and variance $\sigma^2$ denoted by $y \sim \N(\mu, \sigma^2)$. The prediction of $y$ is taken at the mean level, that is, $\hat{y} = \mu$.
They further introduced a \code{Gaussian} output layer in the network architecture, which simultaneously outputs $\bm{\zeta} = (\mu, \sigma^2)$. 

This model is implemented by minimizing the Gaussian NLL loss function,
\begin{equation}
    \Ell_{\N}(y|\mu, \sigma^2) = \frac{1}{2}\log\left(2\pi\sigma^2\right) + \frac{(y - \mu)^2}{2\sigma^2}, \label{eq:gaussian_nll}
\end{equation}
instead of the MSE in classical neural networks.
Each network in the ensemble is trained using stochastic gradient descent.
We remark that this model was originally proposed for non-time-series problems.
Hence, we left out time index $t$ but in our motivating application, variables are indexed by $t$, similar to the DGP in Equation~\eqref{eq:unc_dgp}.

In this setup, $\sigma^2$ models aleatoric uncertainty but the Normal data distribution is incapable of quantifying epistemic uncertainty.
\cite{Lakshminarayanan:2017} addressed this by using an ensemble of neural networks with randomly initialized weights.
Each network settles in a different local minima and produces different $\mu$ and $\sigma^2$ estimates for the same training dataset $\D_t$.
The variance of $\hat{\mu}$ across the ensemble thus provides an empirically-driven proxy for epistemic uncertainty at a computational cost.
Thus, Ensemble does not utilize statistical modelling to capture patterns in epistemic uncertainty.

\subsection{Deep Evidential method} \label{sec:deep_evidential}

\subsubsection{NIG prior distribution} \label{sec:nig_prior}

Addressing the shortcomings of Ensemble, \cite{DeepEvidentialNet:2020} proposed to place an \emph{evidential prior}, the NIG distribution, on the model parameters $\mu, \sigma^2$ of the Normal data distribution:
\begin{align}
    \text{Data}: y &\sim \N(\mu, \sigma^2) \nonumber \\
    \text{NIG prior}: \mu &\sim \N(\gamma, \sigma^2\nu^{-1}), \quad
    \sigma^2 \sim \IG(\alpha, \beta), \label{eq:nig_dist}
\end{align}
where $\mu$ is assumed to be drawn from a Normal prior distribution with unknown mean $\gamma$ and scaled variance $\sigma^2\nu^{-1}$, $\nu$ is a scaling factor for $\sigma^2$, and shape parameter $\alpha > 1$ and scale parameter $\beta > 0$ parameterize the Inverse Gamma (IG) distribution\footnote{Time index $t$ has been omitted for brevity and legibility. Note that variables in this section are indexed by time for each asset: $\{y_t, r_t, \mu_t, \sigma_t^2, \gamma_t, \nu_t, \alpha_t, \beta_t\}$.}.
As Evidential models epistemic uncertainty directly, no ensembling is necessary.
Thus, compared to the Ensemble method, Evidential offers a computational cost advantage and a theoretically motivated epistemic uncertainty estimate.
We require $\alpha > 1$ to ensure the mean of the marginal distribution is finite. 

In this construct, parameters of the posterior distribution of $y$ is $\bm{\zeta} = (\gamma, \nu, \alpha, \beta)$.
Epistemic uncertainty is reflected by the uncertainty in $\mu$ and $\sigma^2$.
The uncertainty in $\mu$ is assumed be a fraction $\nu^{-1}$ of $\sigma^2$ and $\sigma^2$ is assumed to be drawn from an IG distribution.
The fraction $\nu^{-1}$ as controlled by $\nu$ is learnt from the data.
In an abstract sense, $\nu$ varies according to the amount of information in the data and is interpreted as the number of virtual observations for the mean parameter $\mu$.
In other words, $\nu$ number of virtual instances of $\mu$ are assumed to have been observed in determining the prior variance of $\mu$ \citep{ConjugatePrior:2009,DeepEvidentialNet:2020}. 

Based on the NIG prior in Equation~\eqref{eq:nig_dist}, predictions and sources of uncertainty are \citep{DeepEvidentialNet:2020}\footnote{Equation~\eqref{eq:nig_predict0} is due to $\sigma^2 \sim \IG (\alpha,\beta)$ with mean $\frac{\beta}{\alpha-1}$.},
\begin{align}
    \text{Prediction}: \E[\mu] &= \gamma \nonumber \\
    \text{Aleatoric uncertainty}: \E[\sigma^2] &= \tfrac{\beta}{\alpha-1} \label{eq:nig_predict0} \\
    \text{Epistemic uncertainty}: \Var[\mu] &= \tfrac{\beta}{\nu(\alpha - 1)}. \label{eq:nig_predict}
\end{align}
\cite{DeepEvidentialNet:2020} derived these results using integration.
We provide a more straight forward explanation using the marginal distribution of $\mu$ after integrating out $\sigma^2$ from the NIG prior.
\cite{BayesianTheory} showed that this marginal distribution is a non-standardized Student's t-distribution ($\tdist$),
\begin{align}
    \prob(\mu|\gamma, \nu, \alpha, \beta) &= \int_{\sigma^2=0}^{\infty}\prob_{_{\N}}(\mu|\gamma, \sigma^2\nu^{-1}) \prob_{_{\IG}}(\sigma^2|\alpha, \beta) \diff\sigma^2 \nonumber \\
    &= \tdist\left(\gamma,\frac{\beta}{\nu \alpha}, 2\alpha \right). \label{eq:nig_marginal}
\end{align}
Hence, the distribution of  $\mu$ has mean $\gamma$, scale $\frac{\beta}{\nu \alpha}$ and degrees of freedom $2\alpha$ and so, $\Var[\mu] = \frac{\beta}{\nu \alpha} \times \frac{2 \alpha}{2\alpha - 2} = \frac{\beta}{\nu(\alpha - 1)}$.  

From Equations~\eqref{eq:nig_predict0} and \eqref{eq:nig_predict}, $\nu$ can be interpreted as a factor that attributes uncertainty between aleatoric uncertainty ($\tfrac{\beta}{\alpha - 1}$) and epistemic uncertainty ($\tfrac{\beta}{\nu(\alpha - 1)}$).
If $\nu = 1$, then total uncertainty is evenly split between aleatoric and epistemic uncertainties.
We also note that assigning a $\Gam(\alpha, \beta)$ prior to precision $\sigma^{-2}$ in \cite{BayesianTheory} gives the Normal-Gamma (NG) prior and is equivalent to assigning $\IG(\alpha, \beta)$ to $\sigma^{2}$ which gives the NIG prior. 

Apart from the direct explanation of Equation~\eqref{eq:nig_predict}, we also enrich the interpretation of epistemic uncertainty from \cite{DeepEvidentialNet:2020} by decomposing it approximately into uncertainties attributable to parameters $\mu$ and $\sigma^2$.
The Normal prior of $\mu|\sigma^2$ in Equation~\eqref{eq:nig_dist} has variance $\Var[\mu|\sigma^2] = \frac{\sigma^2}{\nu}\approx \frac{\beta}{\nu \alpha}$ \footnote{Since $\sigma^{-2} \sim \Gam(\alpha, \beta)$ with $\E[\sigma^{-2}] = \frac{\alpha}{\beta}$ such that $\E[\sigma^2] = \E[\tfrac{1}{\sigma^{-2}}] \approx \tfrac{1}{\E[\sigma^{-2}]} = \frac{\beta}{\alpha}$.}.
Hence, epistemic uncertainty can be decomposed into,
\begin{align}
    \text{Model $\mu$ uncertainty}: \Var[\mu|\sigma^2] &\approx \tfrac{\beta}{\nu\alpha} \nonumber \\
    \text{Model $\sigma^2$ uncertainty}: \Var[\mu] - \Var[\mu|\sigma^2] &\approx \tfrac{\beta}{\nu\alpha(\alpha - 1)}, \label{eq:model_unc_decomp}
\end{align}
where the difference between the marginal and conditional variances of $\mu$ gives the uncertainty of $\sigma^2$.

In this construct, the marginal distribution of $y$ after integrating out $\mu$ and $\sigma^2$ is a non-standardized Student's t-distribution \citep{DeepEvidentialNet:2020},
\begin{align}
    \prob(y|\gamma, \nu, \alpha, \beta) &= \int_{\sigma^2=0}^{\infty}\int_{\mu=-\infty}^{\infty}\prob_{_{\N}}(y|\mu, \sigma^2)\prob_{_{\NormInvGam}}(\mu, \sigma^2|\gamma, \nu, \alpha, \beta)\diff\mu\diff\sigma^2 \nonumber \\
 &= \tdist\left(y; \gamma, \frac{\beta(1 + \nu)}{\nu\alpha}, 2\alpha\right). \label{eq:nig_t}
\end{align}
Variance of this t-distribution is $\frac{\beta(1 + \nu)}{\nu(\alpha - 1)}$, which corresponds to the sum of epistemic and aleatoric uncertainties,
\begin{equation}
    \Var[y] = \frac{\beta}{\alpha-1} + \frac{\beta}{\nu(\alpha - 1)} = \frac{\beta(1 + \nu)}{\nu(\alpha - 1)}, \label{eq:vary}
\end{equation}
and is $(1+\nu)$ times the variance of $\mu$, the epistemic uncertainty. The corresponding NLL loss function using Equation~\eqref{eq:nig_t} is \citep{DeepEvidentialNet:2020},
\begin{align}
    \Ell_{\NormInvGam}(y|\bm{\zeta}) &= \tfrac{1}{2}\log\left[\tfrac{\pi}{\nu}\right] - \alpha\log\left[2\beta(1+\nu)\right] \nonumber \\
    &+ (\alpha + \tfrac{1}{2})\log\left[(y - \gamma)^2\nu + 2\beta(1+\nu)\right] + \log\left[\tfrac{\Gamma(\alpha)}{\Gamma(\alpha + \tfrac{1}{2})}\right]. \label{eq:nig_nll}
\end{align}
Equation~\eqref{eq:nig_nll} mimics a Bayesian setup, granting neural networks the ability to estimate both epistemic and aleatoric uncertainties, and offers an intuitive interpretation of the model mechanics --- due to uncertainty in the model parameters, the tails of the marginal t-distribution are heavier than a Normal distribution.
This has the effect of regularizing the network and provides an avenue of estimating epistemic uncertainty. 
To encourage low evidence (and hence high predictive uncertainty) in areas of $X$ with high forecast error, \cite{DeepEvidentialNet:2020} proposed to add a total evidence regularizer $\Ell_\mathrm{R} = |y - \gamma|(\nu + 2\alpha)$ to $\Ell_{\NormInvGam}$ (Equation~\eqref{eq:nig_nll}), where $\nu$ presents virtual observations for the mean and $2\alpha $ for the variance.
As the distribution of asset returns has heavy tails \citep{Cont:2001}, we argue that the marginal t-distribution also provides better a fit for noisy financial time-series.

\subsubsection{Shortcomings} \label{sec:shortcomings}

Among the four parameters $(\gamma, \nu, \alpha, \beta)$, it is clear that the training of $\gamma$ in the neural network is direct as it corresponds to mean of the t-distribution in Equation~\eqref{eq:nig_t}.
By contrast, the scale of t-distribution is modelled through a more complex construct ($\tfrac{\beta(1 + \nu)}{\nu\alpha}$), which reflects the two sources of uncertainty in Equation~\eqref{eq:vary} with two additional parameters: the shape ($\alpha$) and scale ($\beta$) parameters of the Gamma prior distribution for precision $\sigma^{-2}$.
\cite{Bengs:2023} argue that learning $\alpha, \beta$ in the hierarchical model (Equation \eqref{eq:nig_dist}) requires $\mu, \sigma^2$ which are not directly observed.
Thus, hierarchical models lack theoretical guarantee on the robustness of their estimated distributions.

Moreover, \cite{Meinert:2022} notes that Equation~\eqref{eq:nig_t} is overparameterized, as it is possible to minimize Equation~\eqref{eq:nig_nll} irrespective of $\nu$ by sending $\nu \to 0$ (i.e., independent of the data), since $\tfrac{\partial}{\partial\nu}\Ell_{\NormInvGam} = 0$ along the path of $\beta\nu =\tfrac{1}{1 + \nu^{-1}}$. 
This is because Equation~\eqref{eq:nig_t} is, by definition, a projection of the NIG distribution, and thus is unable to unfold all of its degrees of freedom unambiguously \citep{Meinert:2022}.
Through simulation data, \cite{Meinert:2022} showed that over the course of neural network training, the estimated $\nu$, which controls the ratio of epistemic uncertainty to aleatoric uncertainty, was related to speed of convergence.
The results also show that the estimated $\nu$ may not be accurate.
We note that this is also evident in Equation~\eqref{eq:nig_t}, as $\nu$ appears in both the numerator and denominator of ($\tfrac{\beta(1 + \nu)}{\nu\alpha}$), the scale parameter of t-distribution in the form of $1+\frac{1}{\nu}$.
Thus, $\nu$ relates ambiguously to the scale parameter of the t-distribution.


\subsubsection{Architecture of neural network}

\begin{figure}[H]
    \centering
    \includegraphics[width=0.7\linewidth]{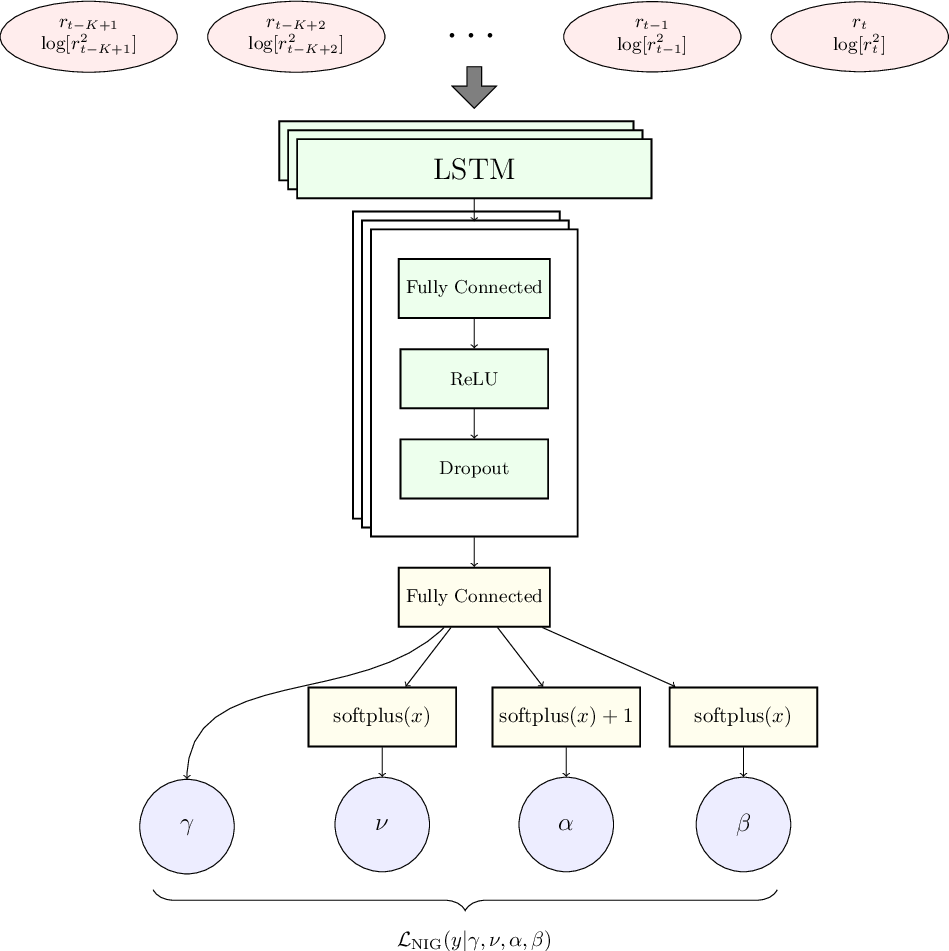}
    \caption{Our implementation of Evidential for cryptocurrency and U.S. equities forecasting. Output from the LSTM layers is fed into a single subnetwork, consisting of one or more blocks of fully connected and dropout layers. For illustrative purposes, we have shown the network with three LSTM layers and three hidden layer blocks. The actual number of hidden layers is subject to hyperparameter tuning (search range detailed in Appendix~\ref{apd:unc_params}). The \code{NormalInverseGamma} output layer (shaded in yellow) is a fully connected layer which outputs four hyperparameters of the marginal distribution. Softplus is applied to $\nu, \alpha, \beta$.}
    \label{fig:evidential_network}
\end{figure}

Let $\bm{a} \in \R^{H^{(I)}}$ be the input vector of the final layer with $H^{(I)}$ dimensions and $H^{(O)}$ be the dimension of the output layer.
In the case of the \code{NormalInverseGamma} layer, $H^{(O)} = 4$ as it outputs,
\begin{align}
    \bm{\zeta} = \Out(\bm{a};\bm{\theta}) = \bm{a}^\top \cdot \bm{W}^{(O)} + \bm{b}^{(O)} \nonumber \\
    \gamma = \zeta_1, \quad \nu = \zeta_2, \quad \alpha = \zeta_3, \quad \beta = \zeta_4, \label{eq:nig_output_layer}
\end{align}
where $\Out$ denotes the \code{NormalInverseGamma} output layer, $\{\zeta_{1:4}\}$ are \nth{1}, ..., \nth{4} elements of vector $\bm{\zeta}$, $\bm{W}^{(O)} \in \R^{H^{(I)} \times H^{(O)}}$ and $\bm{b}^{(O)} \in \R^{H^{(O)}}$ are weights and bias of the output layer, respectively.
Each dimension of $\bm{\zeta}$ corresponds to each of $\gamma, \nu, \alpha$ and $\beta$.
Network architecture is illustrated in Figure~\ref{fig:evidential_network}, showing common LSTM and fully connected layers for $\bm{\zeta}$.
The overall network architecture is the same as our Combined method (Figure~\ref{fig:individual_stacks}) to facilitate comparison.

As a further critique of the network architecture, we note that all four hyperparameters of Evidential are derived from the same latent representation outputted by the last hidden layer.
Outputs of the \code{NormalInverseGamma} layer are linear transformations of a common input $\bm{a}$ (Equation~\eqref{eq:nig_output_layer}).
The four hyperparameters can have vastly different scales (e.g., in our motivating application, $\gamma$ is in scale of 0.01, while $\nu$ is in scale of 10).
We consider this feature to be a weakness of these approaches as the latent representation has to provide a sufficiently rich encoding to linearly derive all hyperparameters of the distribution.
Nonetheless, successful applications of Evidential on real world datasets has led \cite{Meinert:2022} to conclude that Evidential is a heuristic to Bayesian methods and may be appropriate for applications that aim to capture both aleatoric and epistemic uncertainties but do not demand an accurate distinction between them, such as our motivating application.
Motivated by these observation, we propose a simpler formulation, which is more efficient in capturing both uncertainties.
We detail the formulation in Section~\ref{sec:unc_scale_mixture}.

\subsection{Proposed combined method} \label{sec:combined}

\subsubsection{Scale mixture distribution} \label{sec:unc_scale_mixture}

As discussed in Section~\ref{sec:nig_prior}, Evidential provides the ability to perform granular attribution of uncertainty to various parts of the model (e.g., Equation~\eqref{eq:nig_predict} and \eqref{eq:model_unc_decomp}).
However, this ability comes at the cost of model complexity and the estimated epistemic uncertainty may not be reliable (see Section~\ref{sec:shortcomings}).
We sought to propose a simpler model formulation addressing the concerns of \cite{Bengs:2023} on hierarchical models with latent variables and \cite{Meinert:2022} on overparametrization with unresolved degrees of freedom in the Evidential method.

We propose to simplify the model by formulating the problem as a SMD\footnote{Time index $t$ has been omitted for brevity and legibility. Note that variables in this section are indexed by time for each asset: $\{y_t, \gamma_t, \sigma_t^2, \nu_t, \alpha_t, \beta_t\}$. We use the same notations in Equation~\eqref{eq:smd} as Equation~\eqref{eq:nig_dist}, where the symbols have the same meaning to improve comparability.} (\citealp{Andrews:1974}),
\begin{align}
    y|\nu \sim \N(\gamma, \sigma^2\nu^{-1}), \quad
    \nu \sim \Gam(\alpha, \beta), \label{eq:smd}
\end{align}
where $\nu > 0$ is the scaling factor.
This scaling factor affords flexibility to inflate the variance (by minimizing $\nu$ without inflating $\sigma^2$) so as to capture the extremities of the distribution.

We note that uncertainty of variance can be modelled through either $\sigma^2$ or $\nu^{-1}$ as they are indistinguishable in $\sigma^2\nu^{-1}$ in Equation~\eqref{eq:smd}.
Hence, placing the IG prior on $\sigma^2$, that is $\sigma \sim \IG(\alpha, \beta)$, is equivalent to placing a Gamma prior on the scaling factor $\nu$.
However, our proposed formulation is distinct from using a NIG prior as it effectively omits the prior on $\mu$ and lets $y$ to take the distribution of $\mu$. 
This is motivated by our asset return forecasting application, where the mean is typically close to zero (in scale of 0.01) and thus uncertainty of $\mu$ is negligible, whereas volatility is significantly larger (standard deviation in scale of 0.1).
This idea is consistent with fitting a return series with models such as Generalized ARCH (GARCH; \citealp{GARCH:1986}) in which the mean process is typically assumed zero or first order autoregressive \citep{StockVolatilityBook:Ch04}.

The marginal distribution of the Normal model in Equation~\eqref{eq:smd} is a non-standardized t-distribution,
\begin{align}
    \prob(y|\gamma, \sigma^2, \alpha, \beta) &= \int_{\nu=0}^{\infty}\prob_{_{\N}}(y|\gamma, \sigma^2\nu^{-1})\prob_{_{\G}}(\nu|\alpha, \beta)\diff\nu \nonumber \\
    &= \tdist\left(y;\gamma, \frac{\sigma^2\beta}{\alpha}, 2\alpha\right), \label{eq:smd_t}
\end{align} 
which is similar to Equation~\eqref{eq:nig_marginal} with $y$ replacing $\mu$, has the same degrees of freedom parameter and $\sigma^2$ replacing $\nu$ in the parameter set.
This is an advantage as $\sigma^2$ has a richer interpretation --- it directly indicates the scale of the conditional data distribution.
Derivation of Equation~\eqref{eq:smd_t} is provided in Appendix~\ref{sec:smd_marginal}. 
Conceptually, this marginal t-distribution can be interpreted as the Normal distribution being ``stretched out'' into a heavier tailed distribution due to the uncertainty in its variance and provides the ability to handle heavy tails of the distribution that characterize asset returns.

Jointly, $\bm{\zeta} = (\gamma, \sigma^2, \alpha, \beta)$ are parameters of the SMD distribution and the outputs of the neural network that minimize 
the NLL loss function,
\begin{align}
    \Ell_{\SMD}(y|\bm{\zeta}) &= \log\left[ \frac{\Gamma(\alpha)}{\Gamma(\alpha+\tfrac{1}{2})} \right] + \tfrac{1}{2}\log[2\pi\sigma^2\beta] 
    + (\alpha + \tfrac{1}{2})\log\left[ \frac{(y - \gamma)^2}{2\sigma^2\beta} + 1 \right], \label{eq:smd_nll}
\end{align}
of the t-distribution in Equation~\eqref{eq:smd_t}.
Derivation is provided in Appendix~\ref{sec:smd_marginal}.
The neural network learns to output parameters in $\bm{\zeta}$ which minimize the NLL, with $\Ell_{\SMD}$ replacing the marginal likelihood function $\log \prob\big(y_{q}|F(\bm{x}_{q};\bm{\theta}^*)\big)$ in Equation~\eqref{eq:generic_loss}.

For the sources of uncertainty, the data $Y|\nu$ conditional on the scaling factor $\nu$ in Equation~\eqref{eq:smd} is Normal with variance given by the scale ($\tfrac{\sigma^2 \beta}{\alpha}$) of the marginal t-distribution.
This variance gives the uncertainty of the data (i.e., aleatoric uncertainty).
The predictive uncertainty, containing both epistemic and aleatoric uncertainties, is given by the variance of the marginal t-distribution.
Hence, the difference between predictive and aleatoric uncertainties gives epistemic uncertainty.
This is illustrated in Equation~\eqref{eq:smd_predict} below:
 \begin{align}
    \text{Prediction}: \E[y] &= \gamma \nonumber \\
    \text{Aleatoric uncertainty}: \E[\tfrac{\sigma^2}{\nu}] &\approx \tfrac{\sigma^2\beta}{\alpha} \nonumber \\
    \text{Predictive uncertainty}: \Var[y] &= \tfrac{\sigma^2\beta}{\alpha}\cdot\tfrac{2\alpha}{2\alpha - 2} = \tfrac{\sigma^2\beta}{\alpha - 1} \label{eq:vary_combined} \\
    \text{Epistemic uncertainty}: \Var[y] - \E[\tfrac{\sigma^2}{\nu}] &\approx \tfrac{\sigma^2\beta}{\alpha - 1} - \tfrac{\sigma^2\beta}{\alpha} = \tfrac{\sigma^2\beta}{\alpha(\alpha - 1)}. \label{eq:smd_predict}
\end{align}

We note that epistemic uncertainty $\tfrac{\sigma^2\beta}{\alpha(\alpha - 1)}$ is smaller than aleatoric uncertainty $\tfrac{\sigma^2\beta}{\alpha}$ by a factor of $\tfrac{1}{\alpha - 1}$, when $\alpha > 2$.
Thus, as $\alpha$ increases, both epistemic uncertainty and scale of the marginal t-distribution monotonically decrease.
Hence, we argue that $\alpha$ is analogous to ``virtual observations'', similar to $\nu$ in NIG.
More importantly, epistemic uncertainty also drops relative to aleatoric uncertainty, as the t-distribution converges to the Normal distribution on increasing $\alpha$.
This stands in contrast to the model with NIG prior in Equation~\eqref{eq:nig_t}, where increasing evidence $\nu$ does not monotonically lead to a decrease in scale ($\tfrac{\beta(1 + \nu)}{\nu\alpha}$) of the t-distribution.

\subsubsection{Addressing the shortcomings of Evidential method} \label{sec:address_shortcome}

Our proposed SMD formulation addresses the concerns of \cite{Bengs:2023} with latent variables and \cite{Meinert:2022} with overparametrization.
Regarding overparametrization, SMD effectively has only three free parameters in Equation~\eqref{eq:smd_t} as parameter $\beta$ is redundant when it appears as a product with $\sigma^2$ in both the marginal NLL (Equation~\eqref{eq:smd_nll}) and all three uncertainty measures (Equation~\eqref{eq:smd_predict}).
Parameters $\sigma^2$ and $\beta$ indicate scales of the Normal and Gamma distributions, respectively.
Together, they contribute to the scale of the marginal t-distribution.
Hence, the number of parameters can be reduced by reparameterizing $\sigma^2 \beta$ as a single parameter. 

Moreover, one can set $\alpha = \beta$, which we consider as the more intuitive choice.
SMD encapsulates several well-known distributions as special cases.
In the case of $\alpha = \beta$, Equation~\eqref{eq:smd} is a Student's t-distribution with $2\alpha$ degrees of freedom, and is Cauchy if $\alpha = \beta = 1$ \citep{Andrews:1974,Choy:2008}.
If $\alpha \neq \beta$, Equation~\eqref{eq:smd} gives the \emph{Pearson Type VII} (PTVII) distribution which can be re-expressed as a Student's t-distribution in Equation~\eqref{eq:smd_t}.
We can, without loss of generality, set $\alpha = \beta$ and reformulate Equation~\eqref{eq:smd_t} as,
\begin{equation}
    \prob(y|\gamma, \sigma^2, \alpha) = \tdist\left(y;\gamma, \sigma^2, 2\alpha\right), \label{eq:a_b_t}
\end{equation}
and the marginal NLL (Equation~\eqref{eq:smd_nll}) as,
\begin{align}
    \Ell_\text{PTVII}(y|\gamma, \sigma^2, \alpha) &= \log\left[ \frac{\Gamma(\alpha)}{\Gamma(\alpha+\tfrac{1}{2})} \right] + \tfrac{1}{2}\log[2\pi\sigma^2\alpha] + (\alpha + \tfrac{1}{2})\log\left[ \frac{(y - \gamma)^2}{2\sigma^2\alpha} + 1 \right]. \label{eq:a_b_nll}
\end{align}
Comparing Equation~\eqref{eq:a_b_t} to the marginal t-distribution using a NIG prior in Equation~\ref{eq:nig_t}, parameters of SMD relate directly to parameters of the t-distribution instead of hyperparameters of the prior distribution, mitigating the concerns of \cite{Bengs:2023} on hierarchical models that estimating hyperparameters $\alpha,\beta$ relies  on latent variables $\mu,\sigma^2$.  
Thus, we argue that SMD offers an attractive trade-off between model complexity and granularity, occupying the middle ground between Ensemble (no prior) and Evidential (prior on both mean and variance).
Lastly, despite advances in regularizing Evidential, Combined with a simplified statistical construct displays superior results in Section~\ref{sec:unc_real_application} without the need for regularization.

\subsubsection{Architecture of the neural network} \label{sec:unc_architecture}

\begin{figure}
    \centering
    \includegraphics[width=0.7\linewidth]{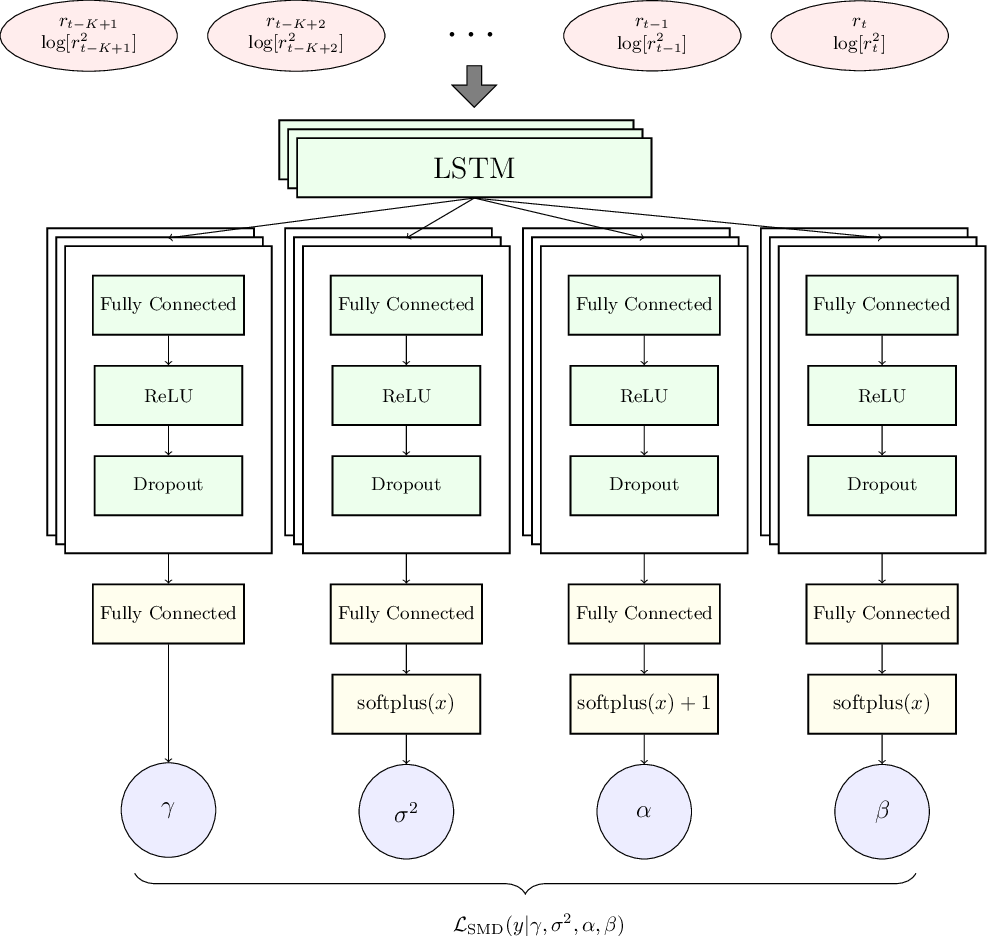}
    \caption{Architecture of our proposed Combined method. Output from the LSTM layers is fed into four subnetworks of one or more blocks, each consisting of a fully connected layer and dropout. The final layer of each subnetwork is a fully-connected layer (shaded in yellow). Softplus is applied to $\sigma^2, \alpha$ and $\beta$ to ensure positivity. The four outputs (shaded in blue) of the neural network are then used to compute $\hat{y}$ and $\Var[\hat{y}]$ in Equation~\eqref{eq:smd_predict}. Note that we can set $\alpha = \beta$, as discussed in Section~\ref{sec:unc_scale_mixture}. In this case, $\beta$ is dropped and the number of subnetworks reduces from four to three.}
    \label{fig:individual_stacks}
\end{figure}

The main application of this paper is the UQ of financial time-series forecasts.
In addition to the statistical model structures, for UQ in financial time-series, network architectural design is also important.
We propose a novel architecture with three distinct improvements.
The proposed architecture is illustrated in Figure~\ref{fig:individual_stacks} and summarized below. 
To predict $\hat{y}_t$, time-series inputs of both \emph{returns} ($r_{t-K+1}, \dots, r_{t}$) and \emph{log-transformed squared returns} ($\log[r_{t-K+1}^2], \dots, \log[r_{t}^2]$) are fed into one or more LSTM layers.
We log-transform squared returns to reduce skewness.
The LSTM layers convert each time-series into a latent representation.
The latent representation is then fed into four independent subnetworks, where each subnetwork is comprised of one or more blocks.
Each block contains a fully connected layer, ReLU and dropout.
In the following, we detail our network architectural design choices.

Firstly, we apply separate subnetworks to each model parameter in $\bm{\zeta}$, whereas in Evidential, the \code{NormalInverseGamma} layer derives four hyperparameters by a linear transformation of a common input $\bm{a}$ (Equation~\eqref{eq:nig_output_layer}).
This is clearly illustrated in Figure~\ref{fig:individual_stacks} relative to Figure~\ref{fig:evidential_network}.
We argue that this construct is too restrictive for complex applications, such as in quantifying uncertainty of financial time-series forecasts.
We propose to model each of the four parameters of SMD with a fully connected subnetwork of one or more layers.
This allows for a more expressive modelling of $\bm{\zeta}$, where each parameter may have complex, non-linear relationships with the input.
As noted in Section~\ref{sec:address_shortcome}, we can set $\alpha = \beta$ and reduce the number of subnetworks to three.
In other words, the network architecture illustrated in Figure~\ref{fig:individual_stacks} can be modified to output three parameters: $\bm{\zeta} = (\gamma, \sigma^2, \alpha)$.
We keep $\beta$ to be comparable to Evidential method in Section~\ref{sec:unc_real_application} but provide empirical results of setting $\alpha = \beta$ in Appendix~\ref{apd:unc_beta_analysis} using the UCI dataset (the same benchmark dataset used in \citealp{Lakshminarayanan:2017} and \citealp{DeepEvidentialNet:2020}, and discussed in Appendix~\ref{apd:uci}) to show that the two networks are indeed equivalent.

In each subnetwork, we enforce constraints on $\sigma^2 > 0$, $\alpha > 1$ and $\beta > 0$ by applying softplus transformation with a constant term, $z' = \log(1 + \exp(z)) + c$, where $z \in \{\sigma^2, \alpha, \beta\}$ and $c$ is the minimum value of the respective parameters.
The transformed values constitute the final output of the network: $\bm{\zeta}' = \{\gamma, (\sigma^2)', \alpha', \beta'\}$.
In Section~\ref{sec:unc_real_application}, we show that this modification vastly improves quantification of forecast uncertainty of financial time-series.
For other network architectures, we argue that the same approach can be applied.
In the case of a feedforward network, we recommend having at least one common hidden layer that reduces the input to a single latent representation.
The latent representation is then passed to individual subnetworks for specialization.
We argue that the common hidden layer allows information sharing across the four parameters, while having no common hidden layer (i.e., if the input is fed into the four disjoint stacks of hidden layers directly) will prevent sharing of information across the stacks.

Secondly, we propose to include the log of squared returns $\{\log(r_{t-K+1}^2), \dots,$ $\log(r_t^2)\}$ as part of the input matrix.
In pooled panel datasets, the model typically learns the average uncertainty within the historical data.
However, as noted in Section~\ref{sec:motivation}, asset returns exhibit time-varying volatility clustering patterns in which the predictive uncertainty is expected to be correlated with time-varying variance of the DGP.
Following the use of squared returns in volatility forecasting literature \citep{Brownlees:2011}, squared returns allow the neural network to infer the prevailing volatility environment.
Squared returns are fed into LSTM in similar spirit to the autoregressive terms of squared returns in Generalized Autoregressive Conditional Heteroskedasticity (GARCH; \citealp{Bollerslev1986generalized}), a popular tool for modelling time-varying volatility in statistical models.
GARCH models adopt the same DGP as Equation~\eqref{eq:unc_dgp} with time-varying variance $\sigma_t^2$ modelled using an ARMA model \citep{Box:1994} and time-varying mean $\mu_t$ assuming a fixed value, 0 or some time-series models such as ARMA (leading to the ARMA-GARCH formulation). 

However, our proposed framework also has a few improvements to ARMA-GARCH models as neural networks offer greater flexibility in modelling and can automatically discover interaction between returns and volatility.
This interaction is known as the \emph{leverage effect} \citep{Cont:2001}, where periods of higher volatility is negatively correlated with future asset returns.
By contrast, modelling of interaction effects in GARCH models requires explicit specification by the user in most packages.
LSTM can also be interpreted as having dynamic autoregressive orders (as opposed to fixed orders in GARCH).
The input and forget gates of LSTM allow the network to control the extent of long-memory depending on features of the time-series.
Nonetheless, we do not directly compare against ARMA-GARCH models for two reasons.
First, in this paper, we are focused on advancing UQ methodologies for neural networks.
We argue that several of our advances can be beneficial to both time-series and non-time-series datasets (as demonstrated in Appendix~\ref{apd:uci}).
Second, we lean on the plethora of literature in comparing LSTM to ARMA-variants (e.g., \citealp{Siami-Namini:2018}) and ARCH-variants (e.g., \citealp{Liu:2019}).

Lastly, we adopt model averaging, as a special case of ensembling, for improving predictive power of estimators \citep{Breiman:1996,Goodfellow:2016}.
It was previously shown to improve accuracy of financial time-series forecasting \citep{STAE:2021} and sequential predictions \citep{Raftery:2010}.
As accuracy of both return forecast and predictive uncertainty are important, we propose to incorporate model averaging to improve return forecasts at the cost of higher predictive uncertainty estimates.
For an ensemble of $M$ models, we compute the ensemble forecast $\Tilde{y}$ and predictive variance $\Var[\Tilde{y}]$ as,
\begin{equation}
    \Tilde{y} = \frac{1}{M}\sum_{i=1}^M \hat{y}_i, \quad \Var[\Tilde{y}] = \frac{1}{M}\sum_{i=1}^M(\hat{y}_i^2 + \Var[\hat{y}_i]) - \Tilde{y}^2, \label{eq:unc_ensemble}
\end{equation}
where $\hat{y}_i$ and $\Var[\hat{y}_i]$ in Equation~\eqref{eq:unc_ensemble} are mean and predictive variance of model $i$, respectively.
The higher predictive uncertainty for return forecast $\tilde{y}$ with model averaging stems from the fact that $\E[\hat{y}^2] > \E[\hat{y}]^2$ by Jensen's inequality in 
Equation~\eqref{eq:unc_ensemble}.
Thus, predictive uncertainty using model averaging contains additional uncertainty than those estimated using the marginal t-distribution alone.
In Section~\ref{sec:unc_ablation_study}, we show that model averaging resulted in significant predictive performance improvement and, hence lowest NLL, despite the higher uncertainty estimates.
Procedural-wise, this model averaging is the same as the Ensemble method (hence the name the Combined method of Ensemble and Evidential), but the aim is to improve return forecast accuracy than capturing epistemic uncertainty in Ensemble.

For ease of comparison, we outline the differences of our combined method to Ensemble \citep{Lakshminarayanan:2017} and Evidential \citep{DeepEvidentialNet:2020} methods in Table~\ref{tab:comparison}.
\begin{table}[h]
\caption{A comparison of Combined to Deep Ensemble and Deep Evidential regressions. \emph{Output layer} refers to the structure of output layer(s) of the network that outputs the parameters of the likelihood function.}
\begin{center}
\resizebox{\columnwidth}{!}{%
\begin{tabular}{@{} lccc @{}}
\textbf{Method} & \textbf{Ensemble} & \textbf{Evidential}  & \textbf{Combined} \\
\hline
Prior           & None              & NIG                  & Gamma        \\
Ensemble        & Yes               & No                   & Yes                 \\
Likelihood      & Gaussian          & Student's t          & Student's t         \\
Output layer  & Single layer $\mu, \sigma^2$ & Single layer $\gamma, \nu, \alpha, \beta$ & \makecell{Multiple subnetworks\\for each of $\gamma, \sigma^2, \alpha, \beta$} \\
\end{tabular}
\label{tab:comparison}}
\end{center}
\end{table}   

\section{Experiments} \label{sec:unc_experiments}

Our proposed framework is primarily focused on advancing UQ in time-series exhibiting volatility clustering.
In this section, we detail experiment results in our motivating application --- time-series forecasting and UQ on cryptocurrency and U.S. equities time-series datasets, to illustrate the benefits of our proposed method.
Nonetheless, SMD parameterization, subnetwork construction for each distribution parameter and ensemble predictions can also be applied to general applications of predictive UQ.
In Appendix~\ref{apd:uci}, we also compare our method to Ensemble and Evidential methods using the UCI benchmark dataset.
This is intended to provide readers with a direct comparison to the results published in \cite{Lakshminarayanan:2017} and \cite{DeepEvidentialNet:2020}, demonstrating the benefits of our proposed combined method in non-time-series datasets.

\subsection{Applications in financial time-series}

\label{sec:unc_real_application}

In this section, we will first apply the three methods on the cryptocurrency dataset, then the U.S. equities dataset\footnote{Note that in this section, ``forecast uncertainty'' and ``uncertainty forecast'' refer to estimated predictive uncertainty (i.e., sum of epistemic and aleatoric uncertainties) for simplicity.}.
Prior literature have found both datasets to exhibit time-varying variance (e.g., \citealp{Cont:2001,Hafner:2018}) and so they are suitable for demonstrating our proposed Combined method in comparison with Ensemble and Evidential methods.
To compare performance, we use mean cross-sectional correlation (CC) and root mean square error (RMSE),
\begin{equation}
    \text{CC} = \frac{1}{T}\sum_{t=1}^T\rho(\bm{y}_t, \hat{\bm{y}_t}) \hspace{2mm}\mbox{and} \hspace{2mm} \text{RMSE} = \sqrt{\frac{1}{T}\sum_{t=1}^T(\bm{y}_t - \hat{\bm{y}}_t)^\top(\bm{y}_t - \hat{\bm{y}}_t)}, \nonumber
\end{equation}
where $\bm{y}_t$ and $\hat{\bm{y}_t}$ are vectors of target and prediction at time $t$ during the test period and $t$ is hourly for cryptocurrencies and monthly for U.S. equities as measures of predictive accuracy, in addition to NLL as a measure of model fit \citep{InfoCoeff:1974,GrinoldKahn:99,STAE:2021}.
The same neural network architectures are used in the two datasets, with hyperparameters tuned independently.

\subsubsection{Cryptocurrency dataset} \label{sec:crypto_data}

Our cryptocurrency dataset consists of hourly returns downloaded from Binance over July 2018 to December 2021, for ten of the most liquid, non-\emph{stablecoin}\footnote{Stablecoins are cryptocurrencies that are pegged to real world assets (e.g., U.S. Dollar).
As such, they exhibit lower volatility than other non-pegged cryptocurrencies.} cryptocurrencies.
Tickers for these cryptocurrencies are BTC, ETH, BNB, NEO, LTC, ADA, XRP, EOS, TRX and ETC, denominated in USDT\footnote{\emph{Tether} (USDT) is a stablecoin that is pegged to USD. It has the highest market capitalization amongst the USD-linked stablecoins \citep{Lipton:2021}.}.
Data from July 2018 to June 2019 are used for hyperparameter tuning, chronologically split into \SI{70}{\percent} training and \SI{30}{\percent} validation.
Data from July 2019 to December 2021 are used for out-of-sample testing.

A network is trained every 30 days using an expanding window of all cryptocurrency data from July 2018.
Each input sequence consists of 10 days of hourly returns $r$ and log squared returns $\log(r^2)$ (i.e., the input is a matrix with dimensions $240 \times 2$), and are used to predict forward one hour return (i.e., units of analysis and observation are both hourly).
This training scheme is shared across all three models in Table \ref{tab:comparison}.
Network topology consists of LSTM layers, followed by fully connected layers with \emph{ReLU} activation and the respective output layers of Ensemble (\code{Gaussian}) and Evidential (\code{NormalInverseGamma}).
For Combined, we use four separate subnetworks as illustrated in Figure~\ref{fig:individual_stacks}.
As discussed in Section~\ref{sec:proposed_method}, we consider UQ in cryptocurrencies to be especially challenging due to their high volatility.

\subsubsection{U.S. equities dataset} \label{sec:us_equities_data}

Our U.S. equities experiment follows the same setup as cryptocurrencies.
Mimicking the S\&P 500 index universe, the dataset consists of daily returns downloaded from the Wharton Research Data Service over 1984 to 2020, for the 500 largest stocks\footnote{The list of stocks is refreshed every June, keeping the same stocks until the next rebalance.} listed on NASDAQ, NYSE and NYSE American.
Data from 1984 to 1993 are used for hyperparameter tuning (\SI{70}{\percent} training and \SI{30}{\percent} validation), while 1994 to 2020 are used for out-of-sample testing.
Each network is refitted every January using a rolling 10-year window using all assets.
Each input sequence consists of 240 trading days (approximately one-year) of daily returns $r$ and log squared returns $\log(r^2)$, forecasting forward 20-day (approximately one-month) return and its uncertainty.
Note that the unit of analysis is monthly and unit of observation is daily.
One-month is a popular forecast horizon for U.S. equities in literature (e.g., \citealp{Gu:2020} and \citealp{STAE:2021}), which motivated our choice of forecast horizon.

\subsubsection{Empirical results} \label{sec:empirical_results}

Table~\ref{tab:main_results} reports forecast results for both cryptocurrency (left) and U.S. equities (right) datasets.
We observe that Combined method has the highest average cross-sectional correlation (higher is better), and lowest RMSE and NLL (both lower is better) in both datasets.
This indicates that Combined method has higher cross-sectional predictive efficacy (as measured by correlation) and is able to better forecast uncertainty of the time-series prediction.
In cryptocurrency, Evidential method has higher (better) correlation, lower RMSE (better) but higher (worse) NLL than Ensemble method.
In U.S. equities, Evidential method has worse correlation, RMSE and NLL than Ensemble method.
Correlations in U.S. equities are materially lower for all three methods compared to the cryptocurrency dataset.
We hypothesize that this is due to both the difference in forecast horizon and maturity of the U.S. stock market.
\begin{table}[h]
\caption{\textbf{Main results}: Comparing Ensemble, Evidential and Combined methods on average cross-sectional correlation, RMSE and NLL for cryptocurrencies (left) and U.S. equities (right), respectively. Average result and standard deviation are computed over 10 trials for each method. Best method for each dataset is highlighted in \textbf{bold}.}
\begin{center}
\sisetup{
    table-number-alignment=center,
    detect-all
}
\begin{adjustbox}{max width=\textwidth}
\begin{tabular}{l|ccc|ccc}
 & \multicolumn{3}{c|}{\textbf{Cryptocurrency}} & \multicolumn{3}{c}{\textbf{U.S. equities}} \\
\textbf{Metric} & \textbf{Ensemble} & \textbf{Evidential} & \textbf{Combined} & \textbf{Ensemble} & \textbf{Evidential} & \textbf{Combined} \\
\hline
Correlation ($\times 100$) & $2.78 \pm 1.09$ & $3.94 \pm 1.84$ & $\bm{9.87 \pm 3.17}$ & $0.40 \pm 0.66$ & $0.09 \pm 0.93$ & $\bm{1.22 \pm 0.65}$ \\
RMSE ($\times 100$) & $0.874 \pm 0.022$  & $0.874 \pm 0.003$ & $\bm{0.867 \pm 0.001}$ & $9.426 \pm 0.044$ & $9.433 \pm 0.033$ & $\bm{9.379 \pm 0.020}$ \\
NLL & $-3.74 \pm 0.10$  & $-3.24 \pm 0.02$  & $\bm{-4.14 \pm 0.01}$ & $-1.65 \pm 0.17$ & $-0.82 \pm 0.03$ & $\bm{-1.71 \pm 0.01}$ \\
\end{tabular}
\label{tab:main_results}
\end{adjustbox}
\end{center}
\end{table}

\begin{figure}[h]
    \centering
    \begin{subfigure}[t]{0.49\textwidth}
        \centering
        \includegraphics[width=\linewidth]{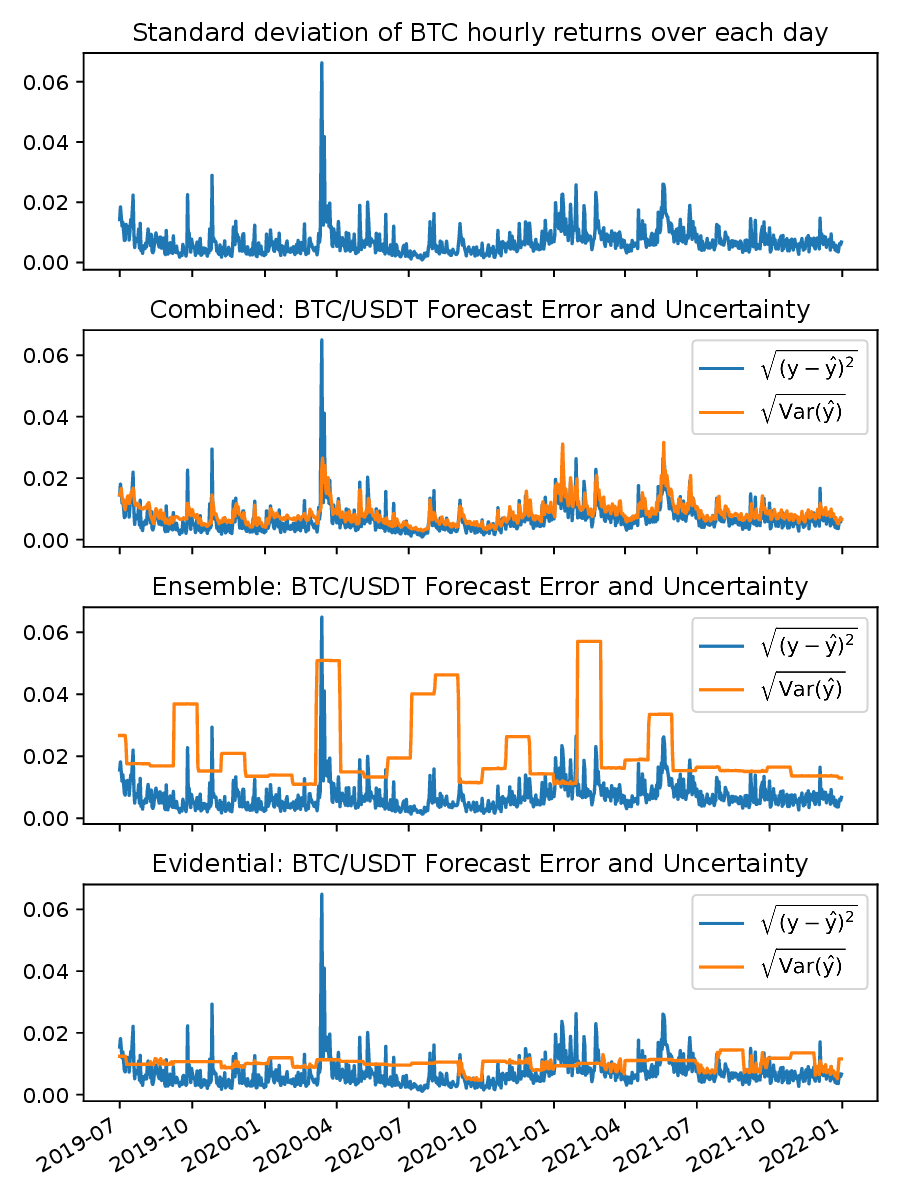}
        \caption{Uncertainty of Bitcoin}
        \label{fig:btc}
    \end{subfigure}
    \begin{subfigure}[t]{0.49\textwidth}
        \centering
        \includegraphics[width=\linewidth]{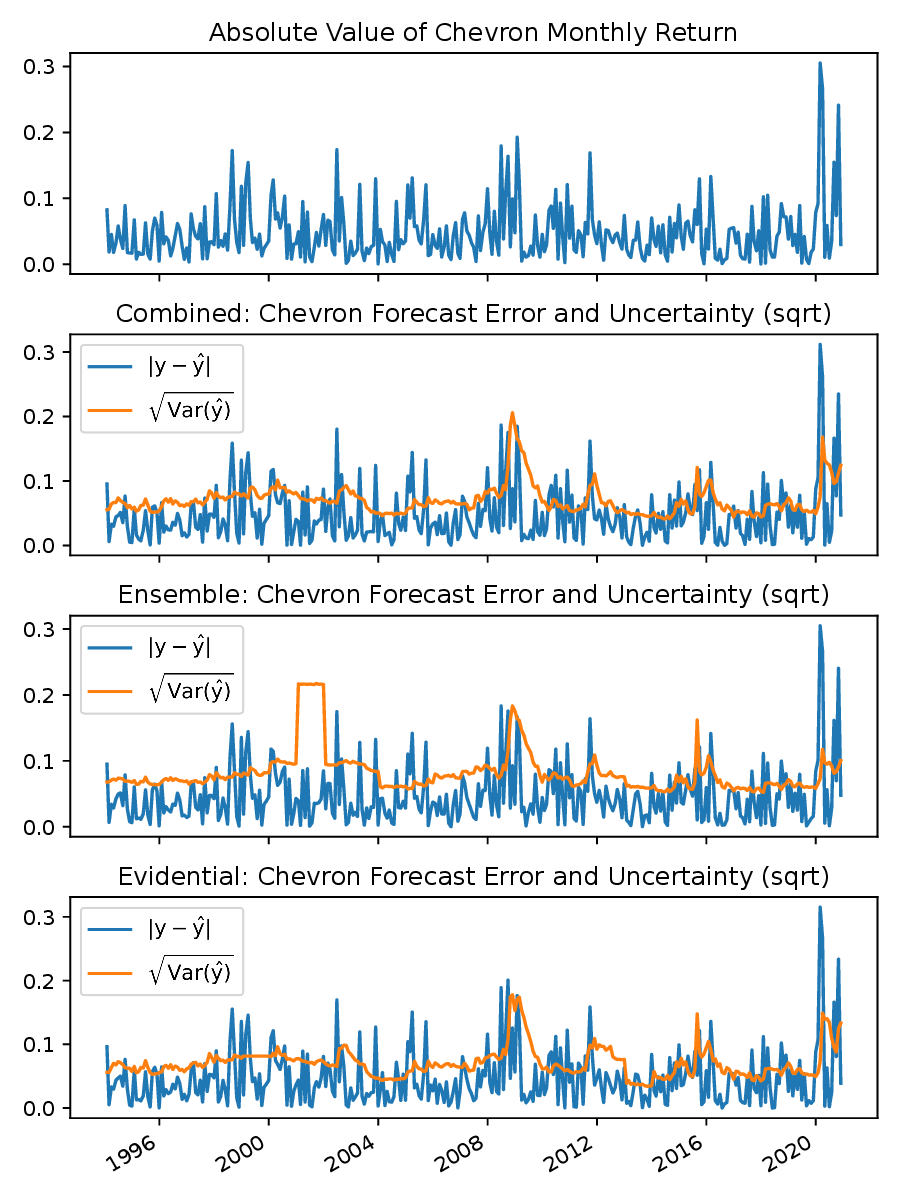}
        \caption{Uncertainty of Chevron}
        \label{fig:cvx}
    \end{subfigure}
    \caption{Forecast uncertainties of Combined, Ensemble and Evidential applied to Bitcoin (left; daily) and Chevron Corporation (right; monthly), respectively over the out-of-sample period. For the top row, SD for Bitcoin and AV for Chevron in Equation~\eqref{eq:volatility_clustering} are shown.}
    \label{fig:btc_uncertainty}
\end{figure}

To compare relative UQ performance, we apply the trained networks to Bitcoin (BTC/USDT), the cryptocurrency with the highest market capitalization, and Chevron Corporation, a major U.S. oil producer which have endured multiple market shocks.
Additionally, we use the standard deviation (SD) of hourly returns computed over each day for Bitcoin, and absolute value (AV) of monthly returns $|y_t|$ for Chevron to provide visual confirmation of volatility clustering, given respectively by,
\begin{equation}
    \text{SD} = \sqrt{\frac{1}{24}\sum_{k = 0}^{23}(y_{t-k} - \bar{y}_{t-k})^2} \quad \text{and} \quad \text{AV} = |y_t|, \label{eq:volatility_clustering}
\end{equation}
where $t$ is hourly and monthly time. 
Figure~\ref{fig:btc_uncertainty} visualizes these volatility clustering in the top row.
As expected, Bitcoin displays sharper spikes of clustered volatility due to its more volatile nature.
For Chevron, we observe multiple spikes of volatility, corresponding to the U.S. recession over 2008-09, the oil shock in 2015 and the 2020 pandemic.
These spikes are relatively lower than those of Bitcoin due to the lower frequency of return forecast for Chevron (monthly) relative to Bitcoin (hourly).
Hence, the dynamics of returns volatility differ
across horizon of returns.

Next, we evaluate the performance of predictive uncertainty forecast during the test period.
\emph{Predictive uncertainty} are given by $\Var[\hat{y}] = \hat{\sigma^2}$ for Ensemble in Equation~\eqref{eq:gaussian_nll}, $\Var[y] = \tfrac{\hat{\beta}(1 + \hat{\nu})}{\hat{\nu}(\hat{\alpha}-1)}$ for Evidential method in \eqref{eq:vary} and $\Var[\hat{y}] = \Var[\tilde{y}]$ for Combined method in \eqref{eq:unc_ensemble}.
They should then be compared with observed volatility over a look back window. 
However, the true instantaneous volatility of an asset (i.e., $\sigma^2$ in Equation~\eqref{eq:unc_dgp}) is not directly observable \citep{Ge:2022}.
Hence, we estimate the true volatility using different formulations of \emph{actual forecast error} for the returns of Bitcoin and Chevron Corporation as they have different units of analysis. 
For Bitcoin, we compute the daily root mean squared return forecast error (RMSE) using hourly return forecast as the true daily volatility proxy and compare it to daily root mean predictive uncertainty (RMPU) given by respectively,
\begin{equation}
    \text{RMSE} = \sqrt{\frac{1}{24}\sum_{k = 0}^{23}(y_{t-k} - \hat{y}_{t-k})^2} \quad \text{and} \quad  \text{RMPU} = \sqrt{\frac{1}{24}\sum_{k = 0}^{23}\Var(\hat{y}_{t-k})}, \nonumber
\end{equation}
for each $t = 24, 48, 72, \dots, T$ where $t$ for cryptocurrency is in hourly units. The two measures, true proxy and predictive uncertainty, are denoted by $\sqrt{(y - \hat{y})^2}$ and $\sqrt{\Var(\hat{y})}$ respectively.
For Chevron Corporation, as the forecast horizon is monthly, we adopt the absolute monthly return forecast error (AE) between actual monthly returns and predicted returns (denoted by $|y - \hat{y}|$) as a proxy for the true $\sigma^2$ and the square-root of forecast uncertainty (MAPU; denoted by $\sqrt{\Var(\hat{y})}$).
We note that SD for Bitcoin or AV for Chevron in Equation~\eqref{eq:volatility_clustering} are very similar to RMSE for Bitcoin and AE for Chevron since $\hat{y} \approx \bar{y} \approx 0$.

Figure~\ref{fig:btc_uncertainty} compares these \emph{predictive uncertainty} $\big(\sqrt{\Var(\hat{y})}\big)$ (orange line) with \emph{actual prediction errors} ($\sqrt{(y - \hat{y})^2}$ or $|y - \hat{y}|$; blue line) for the three methods (row 2-4) applied to Bitcoin (left column) and Chevron Corporation (right column).
We observe in these plots that predictive uncertainty spikes 
when the actual forecast error of the asset spikes.
This is expected, as the spike in volatility (measured by actual forecast errors) leads to large predictive uncertainty. 
Comparing row 2-4 of Figure~\ref{fig:btc_uncertainty}, which correspond to Combined, Ensemble and Evidential, respectively, we observe that Combined's predicted uncertainty of $\hat{\mu}$ tracks actual forecast error much more closely than Evidential and Ensemble.
This appears to be especially true during periods of elevated volatility, which are important to investors.
Overestimation of predictive uncertainty is severe for Ensemble in Bitcoin, where predictive uncertainty can sometimes be significantly higher than actual forecast error.
Lastly, Evidential underestimates forecast error during heightened volatility (e.g., March 2020 in Figure~\ref{fig:btc_uncertainty}) and overestimates forecast error under periods of low volatility (e.g., July 2020 in Figure~\ref{fig:btc_uncertainty}).
On the other hand, we also observe that Combined method tracks the spike in forecast error better than Ensemble and Evidential.

We note that the ``block-like'' appearances of uncertainty forecasts of both Ensemble and Evidential are due to periodic training (monthly for cryptocurrencies and yearly for U.S. equities) and the failure to generalize the prevailing volatility environment.
During training, the optimizer updates network weights $\bm{W}$ and bias $\bm{b}$ (which is analogous to the intercept in linear models).
When the network fails to generalize, it minimizes the loss function by updating the bias rather than the weights.
Thus, outputting the same constant that do not vary with the input, until the network is re-trained in the following month (for cryptocurrencies) or year (for U.S. equities).
This produces the block-like appearances of Ensemble and Evidential, and is indicative of the network setup (e.g., no separate modelling of hyperparameters) being unsuitable to this class of problems.
We observe similar visual characteristics in the predicted uncertainty of other cryptocurrencies and stocks.

\begin{figure}[h]
    \centering
    \begin{subfigure}[t]{0.49\textwidth}
        \centering
        \includegraphics[width=\linewidth]{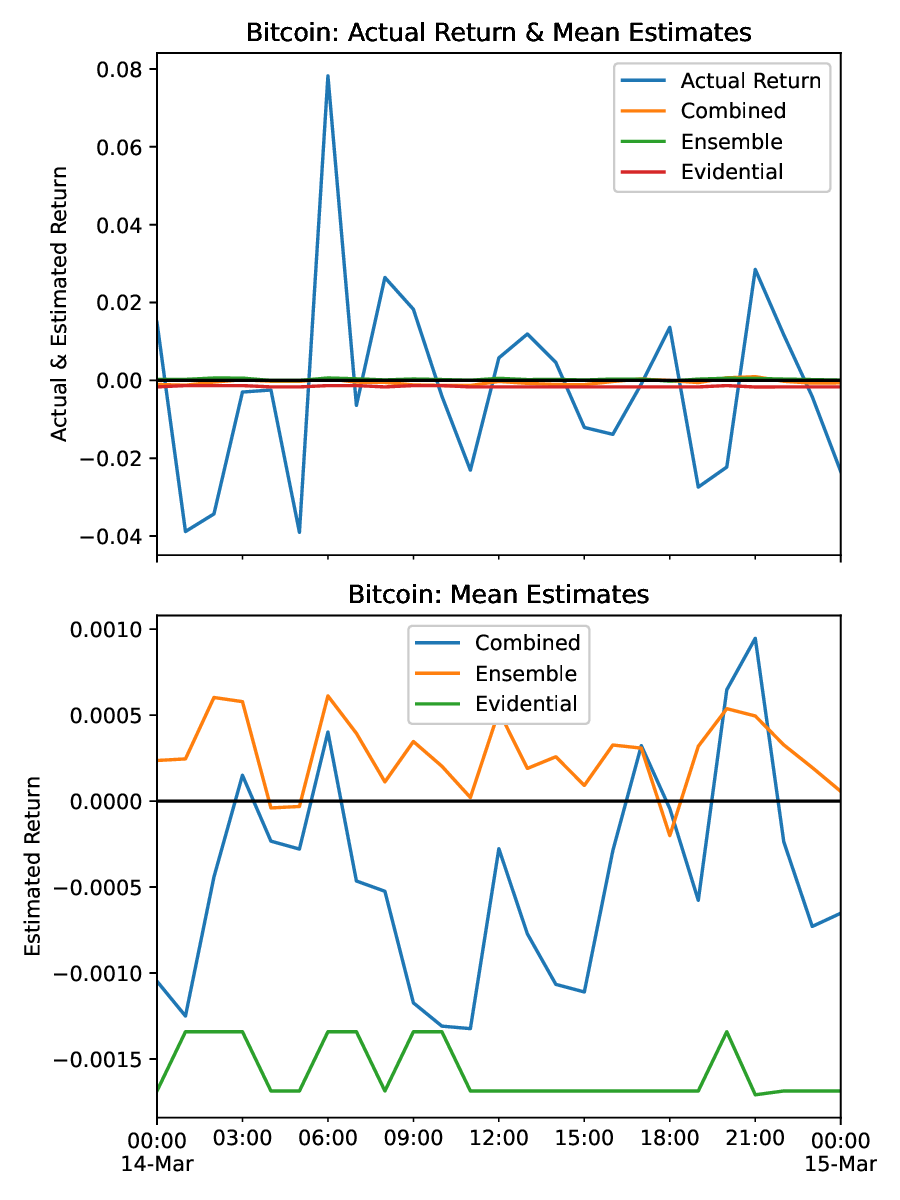}
        \caption{\footnotesize Actual \& Estimated Returns of Bitcoin}
        \label{fig:btc_mean}
    \end{subfigure}
    \begin{subfigure}[t]{0.49\textwidth}
        \centering
        \includegraphics[width=\linewidth]{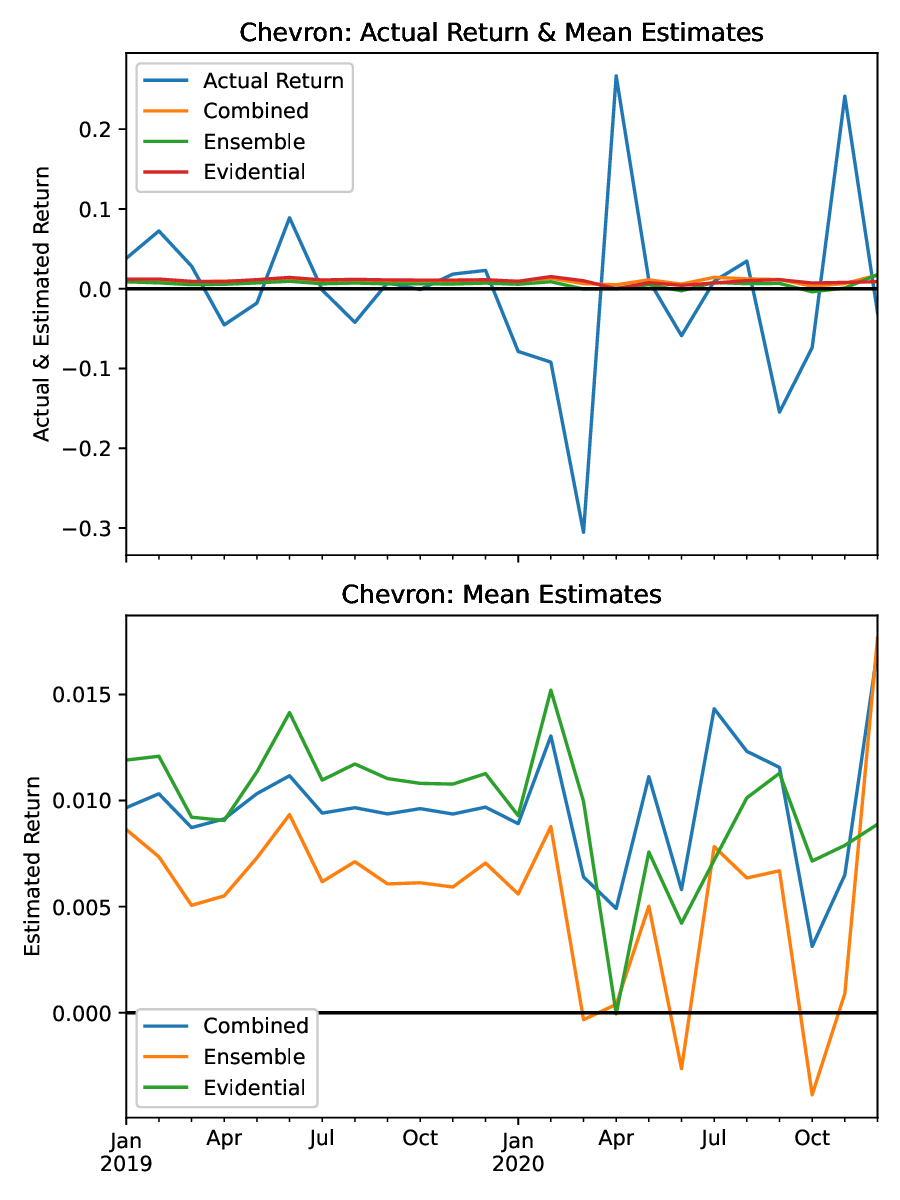}
        \caption{\footnotesize Actual \& Estimated Returns of Chevron}
        \label{fig:cvx_mean}
    \end{subfigure}
    \caption{Actual returns and mean estimates of Combined, Ensemble and Evidential applied to Bitcoin (left; hourly) and Chevron (right; monthly), respectively. For legibility, we have shown hourly forecasts on March 14, 2020, and monthly forecasts over 2019-2020 for Bitcoin and Chevron, respectively.}
    \label{fig:mean_estimates}
\end{figure}

Lastly, in Figure~\ref{fig:mean_estimates}, we compare the mean estimates of Combined, Ensemble and Evidential to actual realized returns.
To improve legibility, we have shown hourly forecasts for Bitcoin on March 14, 2020, a day of high volatility in cryptocurrencies, and monthly forecasts for Chevron over 2019-2020.
The top row of Figure~\ref{fig:mean_estimates} compares mean estimates of the three models to actual realized returns of Bitcoin (left) and Chevron (right).
Due to the low signal-to-noise ratio in financial data, the mean estimates have materially smaller scales than actual returns for all three models.
Next, focusing on the bottom row of Figure~\ref{fig:mean_estimates}, we observe higher variability for Combined in Bitcoin than Ensemble and Evidential.
Evidential produced estimates that were persistently negative over this 24-hour window and did not anticipate the fluctuations of actual returns.
Turning to Chevron, all three methods predict positive returns throughout 2019 but the fluctuations in estimates appear to track the dip in April and jump in June.
However, all three methods could only forecast the sharp fall and subsequent reversal in returns over March and April 2020 at a lag of one month.
This is to be expected as the volatility was due to an exogenous shock (pandemic).

\subsection{Ablation study} \label{sec:unc_ablation_study}

\begin{figure}[h]
    \centering
    \begin{subfigure}[t]{.49\textwidth}
        \centering
        \includegraphics[width=\linewidth]{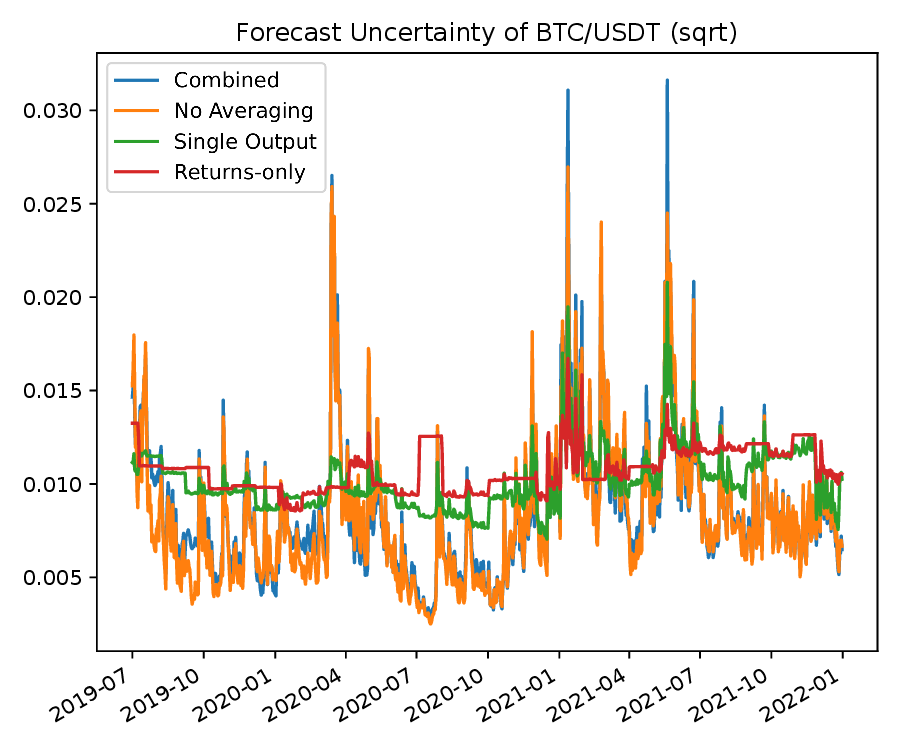}
        \caption{Uncertainty of Bitcoin}
        \label{fig:btc_ablation}
    \end{subfigure}
    \begin{subfigure}[t]{.49\textwidth}
        \centering
        \includegraphics[width=\linewidth]{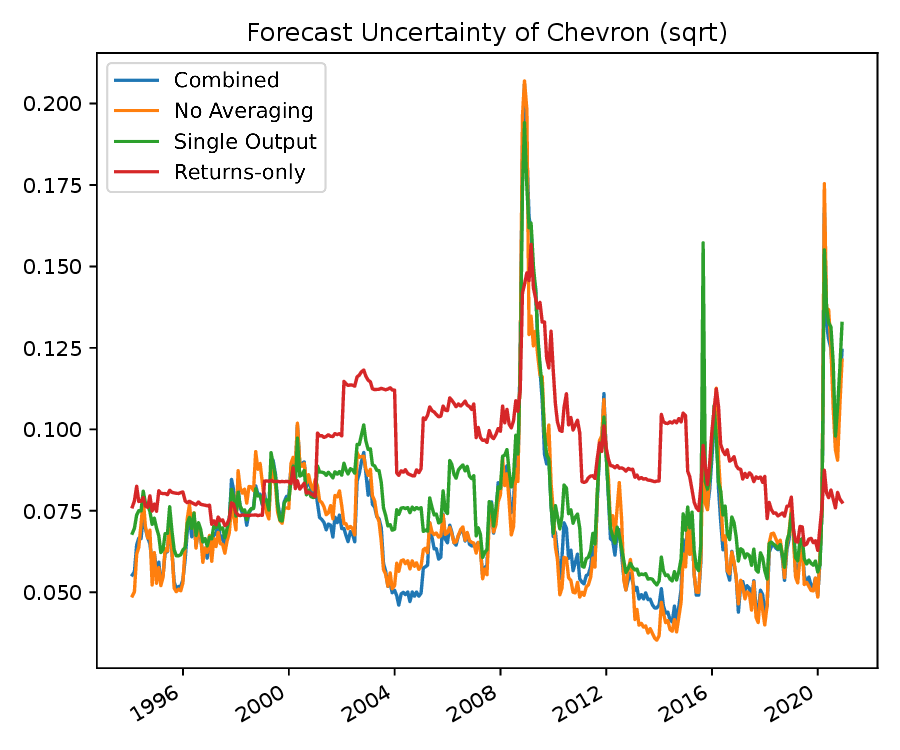}
        \caption{Uncertainty of Chevron}
        \label{fig:cvx_ablation}
    \end{subfigure}
    \caption{Predicted uncertainty $\Var(\hat{y})$ of Combined, omitting each of model averaging (\emph{Averaging}), single output layer (\emph{Single Output}) and using returns only (\emph{Returns-only}) for BTC/USDT and Chevron.}
    \label{fig:ablation}
\end{figure}

Next, we test the effects of removing each of the following for Combined: 1) model averaging; 2) single output layer for all distribution parameters (same as Evidential); 3) using return time-series only (i.e., no squared returns).
The results are recorded in Table~\ref{tab:ablation} and in Figure~\ref{fig:ablation}.
As discussed in Section~\ref{sec:unc_architecture}, model averaging (Equation~\eqref{eq:unc_ensemble}) will lead to higher predictive uncertainty estimates.
In Table~\ref{tab:ablation}, we observe that omitting model averaging has a large negative impact on cross-sectional correlation and NLL.
Cross-sectional correlation is \SI{55}{\percent} and \SI{25}{\percent} lower for cryptocurrencies and U.S. equities, respectively.
NLL is also higher by 0.8 in both cases (lower is better), indicating a worse overall fit.
However, it does not appear to impede the network's ability to model time-series forecast uncertainty as shown by comparing Combined (with model averaging; blue line) to No Averaging (without model averaging; orange line) in Figure~\ref{fig:ablation}.
We observe very similar estimated predictive uncertainties with and without model averaging (as the orange and blue lines track each other closely).
This indicates a favorable trade-off between significantly improved return forecast performance and practically the same predictive uncertainty estimates.

\begin{table}[h]
\caption{\textbf{Ablation studies of Combined method}: In each column, we remove model averaging (\emph{No Averaging}), separate modelling of distribution parameters (\emph{Single Output}) and using return time-series only (\emph{Returns-only}) for cryptocurrencies (left) and U.S. equities (right), respectively. Average result and standard deviation over 10 trials for each method. Note that cryptocurrency returns are hourly and U.S. stock returns are monthly.}
\begin{center}
\sisetup{
    table-number-alignment=center,
    detect-all
}
\begin{adjustbox}{max width=\textwidth}
\begin{tabular}{l|ccc|ccc}
 & \multicolumn{3}{c|}{\textbf{Cryptocurrency}} & \multicolumn{3}{c}{\textbf{U.S. equities}} \\
\textbf{Metric} & \textbf{No Averaging} & \textbf{Single Output} & \textbf{Returns-only} & \textbf{No Averaging} & \textbf{Single Output} & \textbf{Returns-only} \\
\hline
Correlation ($\times 100$) & $4.48 \pm 2.80$ & $8.23 \pm 2.91$ & $10.46 \pm 2.04$ & $0.92 \pm 0.65$ & $1.87 \pm 1.06$ & $1.21 \pm 0.73$ \\
RMSE ($\times 100$) & $0.868 \pm 0.001$ & $0.872 \pm 0.002$ & $0.866 \pm 0.002$ & $9.392 \pm 0.020$ & $9.398 \pm 0.029$ & $9.384 \pm 0.046$ \\
NLL & $-3.35 \pm 0.01$ & $-4.04 \pm 0.02$ & $-3.95 \pm 0.02$ & $-0.88 \pm 0.01$ & $-1.63 \pm 0.04$ & $-1.34 \pm 0.04$ \\
\end{tabular}
\label{tab:ablation}
\end{adjustbox}
\end{center}
\end{table}

Using a single output layer for all distribution parameters also leads to marginally worse NLL.
Correlation is lower in cryptocurrencies but marginally higher in U.S. equities.
While using returns only leads to marginally higher correlation but marginally worse on NLL in cryptocurrency, and lower correlation and worse NLL in U.S. equities. 
From Figure~\ref{fig:ablation}, the block-like appearances indicate that both using single output layer and using returns only result in the network failing to closely track time-varying variance of the DGP.
This suggests that both squared returns and separate modelling of distribution parameters are required to model time-varying forecast uncertainty.

\section{Conclusions} \label{sec:unc_conclusion}

Our motivating application of portfolio selection depends on both return forecasts and uncertainties of return forecasts.
This is a challenging problem due to both the low signal-to-noise ratio in financial markets \citep{Gu:2020} and the presence of volatility clustering.
To this end, we present the Combined method for the simultaneous forecasting asset returns and modelling of forecast uncertainty in presence of volatility clustering.
Our proposed method extends and simplifies the work of \cite{Lakshminarayanan:2017} and \cite{DeepEvidentialNet:2020}.
We propose to use a SMD which uses a Gamma prior for scale $\nu$ uncertainty as a simpler alternative to the NIG prior which places a Normal prior to $\mu$ and an Inverse-Gamma prior to $\sigma^2$.
Parameters of SMD are modelled using separate subnetworks.
Together with model averaging and the use of log squared returns as inputs, we show that our proposed method can successfully model time-varying variance of the DGP, while providing superior return forecasting performance than the two state-of-the-art neural network UQ methods --- Evidential and Ensemble.
This is illustrated through the successful quantification of forecast uncertainty of two financial time-series datasets: cryptocurrency and U.S. equities.

Our proposed SMD formulation offers an avenue to resolve some of the criticisms of \cite{Meinert:2022} and \cite{Bengs:2023}.
In particular, our SMD parameterization has three effective parameters and thus does not have any unresolved degrees of freedom.
Setting $\alpha = \beta$ leads to a marginal t-distribution where the three distributional parameters ($\gamma, \sigma^2, \alpha$) relate directly to the location, scale and shape of the t-distribution, without the need of a hierarchical model.
In this formulation, epistemic uncertainty is assumed to be the difference between the predictive (t-distributed) and aleatoric (Normal-distributed) uncertainties.
This assumption prioritizes a simpler model over granular attribution between aleatoric and epistemic uncertainties given by the NIG prior in Evidential.
Moreover, \cite{Meinert:2022} pointed out that the granular control comes at the cost of an unresolved degree of freedom.
This also makes for a potential future research direction to evaluate such cost.

Despite assuming a more relaxed prior to estimate epistemic uncertainty, we show empirically that our method is able to accurately predict forecast errors, similar to the success that Evidential demonstrated in other real world applications (e.g., see \citealp{Liu:2021,Soleimany:2021,Cai:2021,Singh:2022,LiLiu:2022}).
From a finance application perspective, forecast uncertainty can be used to size bets, or as advanced warning to protect the portfolio from downside risk.
For example, if forecast uncertainty reaches a certain threshold, an investor could purchase portfolio insurance (e.g., put options) or liquidate positions to reduce risk.
The ability to attribute epistemic and aleatoric uncertainties may also allow for more advanced portfolio optimization techniques to be developed in future research (e.g., place different risk aversions on the two sources of uncertainties).
Lastly, UQ in time-series applications is a relatively under-explored area of literature.
We believe this paper can lead to further advancements of UQ in complex time-series.
For example, the nonparametric quantile regression of \cite{Huttel:2023}
can be applied to SMD by expressing the AL distribution  hierarchically as scale mixtures of asymmetric uniforms with Gamma mixing density \citep{Choy:2008,wichitaksorn2015analyzing}. Moreover, SMD also has a direct multivariate extension with conditional multivariate Normal and Gamma mixing densities to facilitate multivariate evidential regressions.

\bibliography{uncertainty}
\bibliographystyle{apalike}

\appendix

\section{Marginal distribution of a Scale Mixture} \label{sec:smd_marginal}

From Equation~\eqref{eq:smd}, we have $y|\nu \sim \N(\gamma, \tfrac{\sigma^2}{\nu}), \ \nu \sim {\rm Gamma}(\alpha, \beta)$.
Marginalizing over $\nu$ produces the data likelihood,
\begin{align}
    \prob(y|\gamma, \sigma^2, \alpha, \beta) &= \int_0^{\infty} \prob_{\N}(y|\gamma, \sigma^2\nu^{-1})\prob_{\G}(\nu|\alpha, \beta) \diff\nu \nonumber \\
    &= \int_0^{\infty} \left[ \sqrt{\frac{\nu}{2\pi\sigma^2}}\exp\left\{-\frac{\nu(y-\gamma)^2}{2\sigma^2}\right\} \right] \left[ \frac{\beta^\alpha}{\Gamma(\alpha)}\nu^{\alpha-1}\exp^{-\beta\nu} \right] \diff\nu \nonumber \\
    &= \frac{\beta^\alpha}{\Gamma(\alpha)\sqrt{2\pi\sigma^2}} \int_{0}^{\infty} \nu^{\alpha - \tfrac{1}{2}}\exp\left\{-\frac{\nu(y - \gamma)^2}{2\sigma^2} - \beta\nu\right\} \diff\nu \nonumber \\
    &= \frac{\beta^\alpha}{\Gamma(\alpha)\sqrt{2\pi\sigma^2}} \left[\frac{(y - \gamma)^2}{2\sigma^2} + \beta\right]^{-(\alpha+\tfrac{1}{2})} \int_{0}^{\infty} \left\{ \nu\left[\frac{(y - \gamma)^2}{2\sigma^2} + \beta\right] \right\}^{\alpha-\tfrac{1}{2}} \nonumber \\
    &\quad \exp\left\{-\nu\left[\frac{(y - \gamma)^2}{2\sigma^2} + \beta\right]\right\} \diff\left\{ \nu\left[\frac{(y - \gamma)^2}{2\sigma^2} + \beta\right] \right\}, \nonumber
\end{align}
since $\int_0^{\infty} x^{\alpha-1}\exp(-x) \diff x = \Gamma(\alpha)$,
\begin{align}
    &= \frac{\beta^\alpha}{\sqrt{2\pi\sigma^2}}\frac{\Gamma(\alpha+\tfrac{1}{2})}{\Gamma(\alpha)} \left[\frac{(y - \gamma)^2}{2\sigma^2} + \beta\right]^{-(\alpha+\tfrac{1}{2})} \hspace{19mm} \nonumber
\end{align}
and re-arranging $\beta^{\alpha} = (\tfrac{1}{\beta})^{-\alpha} = (\tfrac{1}{\beta})^{-(\alpha+\tfrac{1}{2})+\tfrac{1}{2}}$,
\begin{align}
    &= \frac{\Gamma(\frac{2 \alpha+1}{2})}{\Gamma(\frac{2\alpha}{2})}\frac{1}{\sqrt{2\pi\sigma^2\beta}} \left[\frac{(y - \gamma)^2}{2\sigma^2\beta} + 1\right]^{-(\frac{2 \alpha+1}{2})} \hspace{42mm} \nonumber \\
    \prob(y|\gamma, \sigma^2, \alpha, \beta) &= \tdist\left(y;\gamma, \frac{\sigma^2\beta}{\alpha}, 2\alpha\right).\label{eq:smd_t_proof}
\end{align}
To show that the last step of Equation~\eqref{eq:smd_t_proof} is true, we start with the probability density function of the t-distribution parameterised in terms of precision $\tdist(y|\gamma, b^{-1}, a)$ \citep{PatternRecognition:2006},
\begin{align}
    \tdist(y|\gamma, b^{-1}, a) &= \frac{\Gamma(\tfrac{a+1}{2})}{\Gamma(\tfrac{a}{2})} \left[\frac{b}{\pi a}\right]^{\tfrac{1}{2}} \left[1 + \frac{b(y - \gamma)^2}{a}\right]^{-(\tfrac{a+1}{2})}, \nonumber
\end{align}
where $\gamma$ is location, $b$ is inverse of scale and $a$ is shape\footnote{Note that the definition of scale $b$ and shape $a$ is used exclusively in this section. Not to be confused with network bias $b$ and activation vector $\bm{a}$ used in the rest of this thesis.}.
Substituting in $b^{-1} = \tfrac{\sigma^2\beta}{\alpha}$ and $a = 2\alpha$,
\begin{align}
    \tdist\left(y|\gamma, \frac{\sigma^2\beta}{\alpha}, 2\alpha\right) &= \frac{\Gamma(\alpha+\tfrac{1}{2})}{\Gamma(\alpha)} \left[\frac{(\tfrac{\alpha}{\sigma^2\beta})}{2\pi\alpha}\right]^{\tfrac{1}{2}} \left[1 + \frac{\alpha(y - \gamma)^2}{2\sigma^2\alpha\beta}\right]^{-(\alpha + \tfrac{1}{2})} \nonumber \\
    &= \frac{\Gamma(\alpha+\tfrac{1}{2})}{\Gamma(\alpha)}\frac{1}{\sqrt{2\pi\sigma^2\beta}} \left[\frac{(y - \gamma)^2}{2\sigma^2\beta} + 1\right]^{-(\alpha+\tfrac{1}{2})}. \nonumber
\end{align}



From Equation~\eqref{eq:smd_t_proof}, the NLL of the marginal t-distribution is,
\begin{align}
    \prob(y|\gamma, \sigma^2, \alpha, \beta) &= \frac{\Gamma(\alpha+\tfrac{1}{2})}{\Gamma(\alpha)}\frac{1}{\sqrt{2\pi\sigma^2\beta}} \left[\frac{(y - \gamma)^2}{2\sigma^2\beta} + 1\right]^{-(\alpha+\tfrac{1}{2})} \nonumber \\
    -\log[\prob(y|\gamma, \sigma^2, \alpha, \beta)] &= \log\left[ \frac{\Gamma(\alpha)}{\Gamma(\alpha+\tfrac{1}{2})} \right] + \tfrac{1}{2}\log[2\pi\sigma^2\beta] + (\alpha + \tfrac{1}{2})\log\left[ \frac{(y - \gamma)^2}{2\sigma^2\beta} + 1 \right]. \nonumber
\end{align}

\section{Benchmarking on UCI dataset for non time-series} \label{apd:uci}

In this section, we compare Combined to Ensemble and Evidential using the UCI benchmark dataset.
This is intended to facilitate a direct comparison to \cite{Lakshminarayanan:2017} and \cite{DeepEvidentialNet:2020} for non-time-series UQ using the same dataset from both papers.
This data set comes from a collection consisting of nine real world regression problems, each with 10--20 features and hundreds to tens of thousands of observations.

\subsection{Comparison in UCI datasets}

\begin{table*}[t]
\caption{Comparing Ensemble \citep{Lakshminarayanan:2017}, Evidential \citep{DeepEvidentialNet:2020} and Combined (this work) on RMSE and NLL using the UCI benchmark datasets. Results are averaged  and standard deviations are calculated over 5 trials for each method. The best method for each dataset and metric are highlighted in \textbf{bold}.}
\begin{center}
\resizebox{\columnwidth}{!}{%
\sisetup{
    table-number-alignment=center,
    detect-all
}
\begin{tabular}{l|ccc|ccc}
 & \multicolumn{3}{c|}{\textbf{RMSE}} & \multicolumn{3}{c}{\textbf{NLL}} \\
\textbf{Dataset} & \textbf{Ensemble} & \textbf{Evidential} & \textbf{Combined} & \textbf{Ensemble} & \textbf{Evidential} & \textbf{Combined} \\
\hline
Boston & $\bm{2.66 \pm 0.20}$ & $2.95 \pm 0.29$ & $2.89 \pm 0.31$ & $2.28 \pm 0.05$ & $2.30 \pm 0.05$ & $\bm{2.23 \pm 0.05}$ \\
Concrete & $5.79 \pm 0.16$ & $5.98 \pm 0.23$ & $\bm{5.40 \pm 0.18}$ & $3.07 \pm 0.02$ & $3.11 \pm 0.04$ & $\bm{2.98 \pm 0.03}$ \\
Energy & $1.86 \pm 0.04$ & $1.84 \pm 0.06$ & $\bm{1.71 \pm 0.20}$ & $1.36 \pm 0.02$ & $1.41 \pm 0.04$ & $\bm{1.35 \pm 0.05}$ \\
Kin8nm & $\bm{0.06 \pm 0.00}$ & $0.06 \pm 0.00$ & $0.06 \pm 0.00$ & $\bm{-1.39 \pm 0.02}$ & $-1.28 \pm 0.03$ & $-1.35 \pm 0.02$ \\
Naval & $\bm{0.00} \pm 0.00$ & $0.00 \pm 0.00$ & $0.00 \pm 0.00$ & $\bm{-6.10 \pm 0.05}$ & $-5.99 \pm 0.09$ & $-5.89 \pm 0.35$ \\
Power & $3.02 \pm 0.09$ & $3.02 \pm 0.08$ & $\bm{2.95 \pm 0.08}$ & $2.57 \pm 0.01$ & $2.56 \pm 0.03$ & $\bm{2.53 \pm 0.02}$ \\
Protein & $3.71 \pm 0.10$ & $4.28 \pm 0.23$ & $\bm{3.67 \pm 0.13}$ & $\bm{2.61 \pm 0.03}$ & $2.73 \pm 0.08$ & $2.70 \pm 0.05$ \\
Wine & $0.60 \pm 0.03$ & $\bm{0.56 \pm 0.02}$ & $0.59 \pm 0.03$ & $0.94 \pm 0.04$ & $\bm{0.92 \pm 0.04}$ & $1.00 \pm 0.03$ \\
Yacht & $\bm{1.22 \pm 0.22}$ & $1.48 \pm 0.47$ & $3.97 \pm 1.06$ & $1.06 \pm 0.08$ & $\bm{0.96 \pm 0.19}$ & $1.17 \pm 0.11$ \\
\end{tabular}
\label{tab:uci}}
\end{center}
\end{table*}

We follow \cite{Lakshminarayanan:2017} and \cite{DeepEvidentialNet:2020} in evaluating our method using RMSE which assesses forecast accuracy and NLL which assesses overall distributional fit, and compare the measures against Ensemble and Evidential.
While we do not explicitly compare inference speed, as our Combined method also uses ensembling, inference speed is expected to be comparable to Ensemble while being slower than Evidential.
We use the source code provided by \cite{DeepEvidentialNet:2020}, with the default topology of a single hidden layer with 50 units for both Ensemble and Evidential\footnote{Source code for \cite{DeepEvidentialNet:2020} is available on Github: \url{https://github.com/aamini/evidential-deep-learning}}.
As individual modelling of distribution parameters (Section~\ref{sec:unc_architecture}) in Combined requires a network with two or more hidden layers, we use a single hidden layer with 24 units, followed by 4 separate stacks of a single hidden layer with 6 units each.
Thus, the total number of non-linear units is 48 (compared to 50 for Ensemble and Evidential).
Note that even though the total number of units are similar across the three models, learning capacity may differ due to different topologies.

Table~\ref{tab:uci} records experiment results.
On RMSE, we find that both Ensemble and Combined have performed well, having the best RMSE in four datasets each.
In two of the sets (\emph{Kin8nm} and \emph{Naval}), all three methods produced highly accurate results that are not separable to two decimal points.
Turning to NLL, we observe a trend towards Combined having lower NLL than the other two methods for four sets, followed by Ensemble with three sets.
Comparing Combined to Evidential, we find that Combined generally has lower RMSE (7 of 9 sets) and NLL (6 of 9 sets).
Although our method is designed for UQ of complex time-series and all 9 datasets are non-time-series datasets, we still observe some improvements in both RMSE and NLL.

\subsection{Ablation studies}

We further present ablation studies on the UCI dataset.
The first study compares Ensemble and Evidential with Single Output (as in Table~\ref{tab:ablation}), which utilizes model averaging and SMD parameterization but not separate modelling of hyperparameters in Combined.
Single Output has the same network topology as Ensemble and Evidential (a single hidden layer with 50 units), as opposed to Combined which has two hidden layers with a total of 48 units. 
We observe from Table~\ref{tab:uci_extra} that Ensemble has the lowest RMSE in 5 (of 9) datasets, followed by Single Output (3 of 9), while Single Output has the best NLL in 6 (of 9) datasets and Ensemble has 3 (of 9).
On both metrics, Evidential has the least favorable performance.
Comparing Combined in Table~\ref{tab:uci} and Single Output in Table~\ref{tab:uci_extra}, Combined has lower RMSE and NLL in 5 of 9 datasets.
Thus, we conclude that separate modelling of hyperparameters provided an incremental benefit on the UCI datasets.

\begin{table}[h]
\caption{Comparing Ensemble, Evidential and Single Output (Combined but without separate modelling of the four parameters of SMD) on RMSE and NLL using the UCI benchmark datasets. Average result and standard deviation over 5 trials for each method. The best method for each dataset and metric is highlighted in \textbf{bold}.}
\begin{center}
\resizebox{\columnwidth}{!}{%
\sisetup{
    table-number-alignment=center,
    detect-all
}
\begin{tabular}{l|ccc|ccc}
 & \multicolumn{3}{c|}{\textbf{RMSE}} & \multicolumn{3}{c}{\textbf{NLL}} \\
\textbf{Dataset} & \textbf{Ensemble} & \textbf{Evidential} & \textbf{Single Output} & \textbf{Ensemble} & \textbf{Evidential} & \textbf{Single Output} \\
\hline
Boston & $\bm{2.66 \pm 0.20}$ & $2.95 \pm 0.29$ & $2.87 \pm 0.18$ & $\bm{2.28 \pm 0.05}$ & $2.30 \pm 0.05$ & $2.29 \pm 0.04$ \\
Concrete & $5.79 \pm 0.16$ & $5.98 \pm 0.23$ & $\bm{5.72 \pm 0.15}$ & $3.07 \pm 0.02$ & $3.11 \pm 0.04$ & $\bm{3.03 \pm 0.02}$ \\
Energy & $1.86 \pm 0.04$ & $\bm{1.84 \pm 0.06}$ & $1.88 \pm 0.04$ & $1.36 \pm 0.02$ & $1.41 \pm 0.04$ & $\bm{1.35 \pm 0.03}$ \\
Kin8nm & $\bm{0.06 \pm 0.00}$ & $0.06 \pm 0.00$ & $0.06 \pm 0.00$ & $\bm{-1.39 \pm 0.02}$ & $-1.28 \pm 0.03$ & $-1.38 \pm 0.02$ \\
Naval & $\bm{0.00 \pm 0.00}$ & $0.00 \pm 0.00$ & $0.00 \pm 0.00$ & $-6.10 \pm 0.05$ & $-5.99 \pm 0.09$ & $\bm{-6.12 \pm 0.06}$ \\
Power & $3.02 \pm 0.09$ & $3.02 \pm 0.08$ & $\bm{2.97 \pm 0.10}$ & $2.57 \pm 0.01$ & $2.56 \pm 0.03$ & $\bm{2.54 \pm 0.02}$ \\
Protein & $\bm{3.71 \pm 0.10}$ & $4.28 \pm 0.23$ & $3.75 \pm 0.11$ & $\bm{2.61 \pm 0.03}$ & $2.73 \pm 0.08$ & $2.72 \pm 0.02$ \\
Wine & $0.60 \pm 0.03$ & $0.56 \pm 0.02$ & $\bm{0.55 \pm 0.02}$ & $0.94 \pm 0.04$ & $0.92 \pm 0.04$ & $\bm{0.92 \pm 0.02}$ \\
Yacht & $\bm{1.22 \pm 0.22}$ & $1.48 \pm 0.47$ & $1.45 \pm 0.33$ & $1.06 \pm 0.08$ & $0.96 \pm 0.19$ & $\bm{0.93 \pm 0.09}$ \\
\end{tabular}
\label{tab:uci_extra}}
\end{center}
\end{table}

In the second study, we further remove model averaging.
The network used is identical to Evidential but trained using the SMD parameterization (i.e., we simply change the loss function in Evidential to Equation~\eqref{eq:smd_nll}).
We observe from Table~\ref{tab:nig_v_ng} that the network trained using the SMD parameterization has lower RMSE in 6 of 9 and lower NLL in 8 out 9 datasets.
We argue that the improved performance of the SMD parameterization is due to its simplicity.

\begin{table}[h]
\centering
\caption{Comparing Normal-Inverse-Gamma and Scale Mixture Distribution on RMSE and NLL using the UCI benchmark datasets. Average result and standard deviation over 5 trials for each method. The best method for each dataset and loss function is highlighted in \textbf{bold}.}
\label{tab:nig_v_ng}
\sisetup{
    table-number-alignment=center,
    detect-all
}
\begin{tabular}{l|cc|cc}
 & \multicolumn{2}{c|}{\textbf{RMSE}} & \multicolumn{2}{c}{\textbf{NLL}} \\
\textbf{Dataset} & \textbf{NIG} & \textbf{SMD} & \textbf{NIG} & \textbf{SMD} \\
\hline
Boston & $\bm{2.95 \pm 0.29}$ & $2.97 \pm 0.20$ & $\bm{2.30 \pm 0.05}$  & $2.31 \pm 0.05$ \\
Concrete & $5.98 \pm 0.23$ & $\bm{5.78 \pm 0.23}$ & $3.11 \pm 0.04$  & $\bm{3.05 \pm 0.04}$ \\
Energy & $\bm{1.84 \pm 0.06}$ & $1.87 \pm 0.16$ & $1.41 \pm 0.04$  & $\bm{1.33 \pm 0.05}$ \\
Kin8nm & $0.06 \pm 0.00$ & $\bm{0.06 \pm 0.00}$ & $-1.28 \pm 0.03$ & $\bm{-1.37 \pm 0.01}$ \\
Naval& $0.00 \pm 0.00$ & $\bm{0.00 \pm 0.00}$ & $-5.99 \pm 0.09$ & $\bm{-6.27 \pm 0.09}$ \\
Power & $3.02 \pm 0.08$ & $\bm{2.98 \pm 0.12}$ & $2.56 \pm 0.03$  & $\bm{2.53 \pm 0.02}$ \\
Protein & $4.28 \pm 0.23$ & $\bm{3.72 \pm 0.16}$ & $2.73 \pm 0.08$  & $\bm{2.39 \pm 0.05}$ \\
Wine & $\bm{0.56 \pm 0.02}$ & $0.56 \pm 0.03$ & $0.92 \pm 0.04$  & $\bm{0.87 \pm 0.04}$ \\
Yacht & $1.48 \pm 0.47$ & $\bm{1.44 \pm 0.49}$ & $0.96 \pm 0.19$  & $\bm{0.91 \pm 0.18}$ \\
\end{tabular}
\end{table}

\section{Further analysis of parameters in a Scale Mixture Distribution} \label{apd:unc_beta_analysis}

In the network architecture proposed in Section~\ref{sec:unc_architecture}, output of the network is $\bm{\zeta} = (\gamma, \sigma^2, \alpha, \beta)$, which parameterises the SMD (Equation~\eqref{eq:smd}).
However, as noted in Section~\ref{sec:address_shortcome}, we can set $\alpha = \beta$ and reduce the number of parameters to three (Equation~\eqref{eq:a_b_t}).
Thus, an alternative specification of the network is to output $\bm{\zeta} = (\gamma, \sigma^2, \alpha)$ (i.e., three parameters instead of four and are computed through three subnetworks, instead of four in Figure~\ref{fig:individual_stacks}).
We label this network \emph{A=B}.
In Table~\ref{tab:combined_s2b}, we compare Combined (4 parameters) with A=B (3 parameters) using the UCI dataset (as introduced in Section~\ref{apd:uci}).
We observe that A=B is better than Combined on 8 (of 9) datasets on RMSE, while Combined is better than A=B on 1 (of 9).
On NLL, A=B is better than Combined on 5 (of 9) datasets, while Combined is better than A=B on 4 (of 9).
Even though A=B has a higher number of datasets with lower RMSE and NLL, we note that the differences are very small and are within margin of error (due to randomness in neural network training).
Thus, we conclude that the two methods provide near identical results but note that A=B is simpler and more interpretable.
However, we choose Combined with four subnetworks to conduct our analysis so that parameters can also be compared with those from Evidential.

\begin{table}[h]
\caption[Empirical results of combining $\sigma^2$ and $\beta$ on UCI dataset]{Comparing A=B (3 parameters) to Combined (4 parameters) on RMSE and NLL using the UCI benchmark datasets. Results are averaged over 5 trials and the best method for each dataset and metric are highlighted in \textbf{bold}.}
\begin{center}
\sisetup{
    table-number-alignment=center,
    detect-all
}
\begin{tabular}{l|cc|cc}
 & \multicolumn{2}{c|}{\textbf{RMSE}} & \multicolumn{2}{c}{\textbf{NLL}} \\
\textbf{Dataset} & \textbf{A=B} & \textbf{Combined} & \textbf{A=B} & \textbf{Combined} \\
\hline
Boston & $2.91 \pm 0.17$ & $\bm{2.89 \pm 0.31}$ & $2.27 \pm 0.04$ & $\bm{2.23 \pm 0.05}$ \\
Concrete & $\bm{5.39 \pm 0.19}$ & $5.40 \pm 0.18$ & $2.99 \pm 0.03$ & $\bm{2.98 \pm 0.03}$ \\
Energy & $\bm{1.56 \pm 0.16}$ & $1.71 \pm 0.20$ & $\bm{1.30 \pm 0.05}$ & $1.35 \pm 0.05$ \\
Kin8nm & $\bm{0.06 \pm 0.00}$ & $0.06 \pm 0.00$ & $\bm{-1.36 \pm 0.02}$ & $-1.35 \pm 0.02$ \\
Naval & $\bm{0.00 \pm 0.00}$ & $0.00 \pm 0.00$ & $-5.87 \pm 0.12$ & $\bm{-5.89 \pm 0.35}$ \\
Power & $\bm{2.93 \pm 0.08}$ & $2.95 \pm 0.08$ & $\bm{2.53 \pm 0.02}$ & $2.53 \pm 0.02$ \\
Protein & $\bm{3.60 \pm 0.10}$ & $3.67 \pm 0.13$ & $2.83 \pm 0.04$ & $\bm{2.70 \pm 0.05}$ \\
Wine & $\bm{0.57 \pm 0.02}$ & $0.59 \pm 0.03$ & $\bm{0.96 \pm 0.03}$ & $1.00 \pm 0.03$ \\
Yacht & $\bm{2.31 \pm 0.43}$ & $3.97 \pm 1.06$ & $\bm{1.11 \pm 0.09}$ & $1.17 \pm 0.11$ \\
\end{tabular}
\label{tab:combined_s2b}
\end{center}
\end{table}

\section{Hyperparameters used in Section~\ref{sec:empirical_results}} \label{apd:unc_params}

In this section, we provide a list of hyperparameter search ranges and mean hyperparameters used to train the neural networks in Section~\ref{sec:crypto_data} and \ref{sec:us_equities_data}.
Hyperparameter search was performed on all combinations of hyperparameters.
Both cryptocurrency and U.S. equities datasets share the same hyperparameter ranges but with hyperparameter search performed separately.

\begin{table}[h]
    \centering
    \caption{Hyperparameter ranges used in Section~\ref{sec:crypto_data} and \ref{sec:us_equities_data}.
    The `LSTM layers' hyperparameter is a list, with the length of the list indicating how many LSTM layers were used and each element of the list indicating the number of units of each LSTM layer.
    Similarly, `Hidden layers' indicate the number of fully connected hidden layers.
    Each element of the list indicate the dimension of that hidden layer.
    ADAM is the optimiser proposed by \cite{ADAM:2015}.}
    \label{tab:unc_models}
    \begin{tabular}{@{} l c @{}}
    \toprule
    Parameter & Search range \\
    \midrule
    LSTM layers & $\{[16, 8], [32, 16, 8], [32, 16], [64, 32, 16]\}$  \\
    Hidden layers & $\{[8], [16, 8]\}$ \\
    Dropout rate & $\{0.2, 0.3, 0.4\}$ \\
    Activation & ReLU \\
    Batch size & 1,000 \\
    Batch normalisation & Yes \\
    Early stopping & Patience 5 / Tolerance 0.0001 \\
    Learning rate $\eta$ & $0.01$ \\
    Optimiser & ADAM \\
    \bottomrule
    \end{tabular}
\end{table}

\end{document}